\documentclass[a4paper,aps,prd,10pt,preprintnumbers,showpacs,twocolumn,superscriptaddress,nofootinbib,amsmath,amssymb,gensymb,amsfonts]{revtex4-1}
\usepackage{graphicx}
\usepackage{cmap}
\usepackage[utf8]{inputenc}
\usepackage[T1]{fontenc}
\usepackage{nicefrac}
\usepackage{bm,amsmath,amssymb,amsfonts,gensymb,bbold}
\usepackage[colorlinks=true,linktocpage=true,linkcolor=blue,citecolor=blue,allcolors=blue]{hyperref}
\usepackage{amsmath}
\usepackage{makecell}
\usepackage{soul}
\usepackage{placeins}
\usepackage{epstopdf}
\graphicspath{{../}}
\def\imo{i}
\def\re#1{Re(#1)}
\def\im#1{Im(#1)}

\newcommand{\eq}[1]{\begin{align} #1 \end{align}}
\begin{document}
\title{Transition from Regular Black Holes to Wormholes in Covariant Effective Quantum Gravity: Scattering, Quasinormal Modes, and Hawking Radiation}
\author{R. A. Konoplya}\email{roman.konoplya@gmail.com}
\affiliation{Research Centre for Theoretical Physics and Astrophysics, Institute of Physics, Silesian University in Opava, Bezručovo náměstí 13, CZ-74601 Opava, Czech Republic}
\author{O. S. Stashko}\email{alexander.stashko@gmail.com}
\affiliation{Tufts University,  Department of Physics and Astronomy, Medford, MA, USA}
\begin{abstract}
Utilizing the Hamiltonian constraints approach, a quantum-corrected solution has been derived \cite{Zhang:2024ney}, which describes either a regular black hole or a traversable wormhole, contingent upon the value of the quantum parameter. In this work, we compute the quasinormal modes  associated with axial gravitational and test fields' perturbations of these objects. We see that due to quantum corrections near the event horizon, the first several overtones deviate from their Schwarzschild values at an increasing rate. The transition between the black hole and wormhole states is marked by modifications in the late-time signal. Our findings reveal that the fundamental quasinormal modes of quantum-corrected black holes exhibit only slight deviations from those of the classical Schwarzschild solution. However, at the transition, the spectrum undergoes significant changes, with the wormhole state characterized by exceptionally long-lived quasinormal modes.  In addition, we calculate absorption cross-sections of partial waves, grey-body factors and energy emission rates of Hawking radiation.     
\end{abstract}

\maketitle
\section{Introduction}

Einstein's General Relativity (GR) has long stood as the cornerstone of our understanding of gravitation and spacetime. However, the theory's predictive breakdown at singularities signals the need for a more fundamental framework, likely rooted in quantum gravity (QG). Among the promising avenues for advancing this framework is the exploration of effective quantum gravity models that seek to retain key elements of GR while incorporating quantum corrections. Such models offer crucial insights into the resolution of singularities and the emergence of novel spacetime structures.

A recent study by Zhang et al. \cite{Zhang:2024ney} presents a significant step in this direction by employing Hamiltonian constraints \cite{Thiemann:2007zz,Ashtekar:2004eh,Zhang:2024khj} to derive a family of quantum-corrected solutions. These solutions describe either regular black holes or traversable wormholes, depending on the value of a quantum parameter $\xi$. The absence of Cauchy horizons in these spacetimes points to enhanced stability under perturbations, setting them apart from previously explored models of quantum-corrected black holes \cite{Zhang:2024khj,Konoplya:2024lch}.  This dual nature—where the object transitions between black hole and wormhole states—introduces fascinating phenomenological implications, including the manifestation of strongly modified late-time response to perturbations.

In this paper, we extend the analysis of these quantum-corrected spacetimes by investigating the quasinormal modes (QNMs) \cite{Kokkotas:1999bd,Konoplya:2011qq} and grey-body factors \cite{Page:1976df,Page:1976ki,Kanti:1995vq} associated with axial gravitational perturbations. QNMs serve as vital probes into the stability and structure of black holes and wormholes, encoding information about the nature of the horizon and the surrounding geometry. Grey-body factors, on the other hand, reveal how spacetime curvature influences the propagation of radiation, impacting the spectrum of outgoing signals.

A key focus of this work is the exploration of how the quasinormal mode spectrum evolves as the quantum parameter $\xi$ drives the transition from a black hole to a wormhole configuration. This transition is characterized by distinctive modifications in the late-time ringdown signal, highlighting the presence of long-lived modes produced as perturbations scatter between the effective potential barriers present in the wormhole regime.
Transitions between black holes and wormholes have been explored in various contexts, including a toy model proposed by Simpson and Visser \cite{Simpson:2018tsi} and several brane-world models \cite{Casadio:2001jg,Bronnikov:2002rn}. However, in the former case, the underlying gravitational theory lacks strong motivation \cite{Bronnikov:2021uta,Alencar:2025jvl}, while in the latter, an infinite number of solutions arise from different choices of the generating function. Perturbations in such configurations have been studied in \cite{Bronnikov:2019sbx,Churilova:2019cyt}. In contrast, the transition presented here is derived in a well-motivated manner as a direct consequence of quantum corrections to Einstein's relativity.

Our analysis shows that the QNM spectrum and grey-body factors for these quantum-corrected solutions differ markedly from their classical Schwarzschild counterparts. This divergence provides a pathway to observationally distinguish quantum-corrected spacetimes from classical black holes, offering a tantalizing opportunity to probe the signatures of quantum gravity through astrophysical observations.

For black holes, this qualitative difference pertains to the behavior of overtones, which deviate from their Schwarzschild values at a rate that increases with the overtone number. Such a pronounced deviation in overtones is associated with modifications in the near-horizon geometry \cite{Konoplya:2022pbc,Konoplya:2023hqb,Konoplya:2022hll} and has been recently observed in several black hole spacetimes \cite{Zinhailo:2024kbq,Konoplya:2023aph,Bolokhov:2023bwm,Konoplya:2024lch,Konoplya:2023kem,Konoplya:2023ahd,Konoplya:2023ppx,Konoplya:2022iyn,Zhu:2024wic,Luo:2024dxl,Zhang:2024nny}.\\

Given that the first overtones are highly sensitive to near-horizon deformations, a more stable characteristic, which relates to the gravitational wave profile \cite{Oshita:2023cjz,Rosato:2024arw} and corresponds to quasinormal modes in the eikonal regime \cite{Konoplya:2024lir,Konoplya:2024vuj}, is the grey-body factor. In this work, we compute the grey-body factors for both black holes and wormholes across various values of the multipole numbers using precise numerical integration.
\\

The paper is organized as follows: In Section II, we provide an overview of the quantum-corrected black hole and wormhole metrics derived via the effective Hamiltonian formalism. Section III outlines the perturbation equations for axial gravitational and test fields. In Section IV, we describe the numerical methods employed to compute quasinormal modes, including the pseudospectral method and time-domain integration. Section V presents a detailed analysis of the quasinormal mode spectra, highlighting the transition from regular black holes to wormholes. In Section VI, we examine scattering properties, including the absorption cross-sections, grey-body factors, and Hawking radiation emission rates. Finally, Section VII summarizes our findings and discusses potential observational signatures and future research directions.

\section{Basic relation}
The Hamiltonian formulation of gravity theories often encounters challenges related to the preservation of general covariance. In \cite{Zhang:2024khj, Zhang:2024ney}, the authors derived an effective Hamiltonian constraint 
$H_{eff}$ that preserves diffeomorphism invariance, without specific choice of gauge. By selecting a specific form of free functions to incorporate quantum gravity effects, they obtained a new solution describing either a regular black holes or traversable wormholes, depending on the value of the parameters \cite{Zhang:2024ney}.
Unlike previous quantum-corrected black hole models, the resulting solution eliminates Cauchy horizons, potentially enhancing its stability against perturbations.

The metric of the quantum-corrected black hole is given by the following line element \cite{Zhang:2024ney} 
\begin{equation}\label{eq:metric}
ds^2=-f(r)dt^2+\frac{1}{f(r)\mu(r)}dr^2+r^2d\Omega^2,
\end{equation}
where $d\Omega^2=(d\theta^2+\sin^2\theta d\phi^2)$ is the metric of two dimensional sphere and
\eq{
&f(r) = 1 - (-1)^n \frac{r^2}{\xi^2} \arcsin\left(\frac{2M \xi^2}{r^3}\right) - \frac{n \pi r^2}{\xi^2},\\
&\mu(r) = 1 - \frac{4 M^2\xi^4 }{r^6}.
}
Here $\xi $ is the quantum parameter, and $M$ is the mass of configuration and $n$ is an arbitrary integer.

We restrict ourselves to the particular case \( n=0 \), which corresponds to asymptotically flat configurations. 
When \( \xi/M < \pi^{3/2}/\sqrt{2} \approx 3.937 \), the spacetime describes a regular black hole metric, where the region that would normally contain the singularity is replaced by a wormhole with a throat located at 
\[
r_m = \sqrt[3]{2} \, \sqrt[6]{M} \, \xi^{2/3}.
\]
For larger values of \( \xi \), the black hole horizon disappears, and the configuration transitions into a traversable wormhole.

The shadows cast by these objects have been analyzed in \cite{Liu:2024iec}, and their gravitational lensing properties have been studied in \cite{Paul:2025wen}.

To address the wormhole (WH) case more effectively, it is convenient to rewrite the line element (\ref{eq:metric}) in quasiglobal coordinates by introducing a new coordinate  $x$ as follows:
\eq{
\label{eq:rtox}
R(x)=\sqrt{x^2+r_m^2},\quad x\in(-\infty,\infty),
}
where $x\to{\pm\infty}$ correspond to spatial infinities of both universes and $x=0$ corresponds to the WH throat. Then (\ref{eq:metric}) has the following form 
\eq{
\label{eq:metric_quasiglobal}
ds^2=-A(x)dt^2+B(x)dx^2+R^2(x)d\Omega^2,
}
where
\begin{eqnarray}
A(x) &=& f(R(x)), \\ 
B(x) &=& \frac{x^2}{(x^2+r_m^2)f(R(x))\mu(R(x))}.
\end{eqnarray}
Various optical phenomena around such black holes have recently been considered in \cite{Liu:2024iec,Paul:2025wen}.

\section{Perturbation equations}
\subsection{Axial gravitational perturbations} 
The direct analysis of perturbations in the Hamiltonian approach \cite{Zhang:2024ney} must be complicated and, to the best of our knowledge, has not been carried out do far for any of the quantum corrected black holes. Instead, it is common to utilize an effective approach \cite{Ashtekar:2018lag,Ashtekar:2018cay,Bouhmadi-Lopez:2020oia,Bonanno:2000ep} where the metric (\ref{eq:metric}) appears as a solution of the standard Einstein equations, but with the presence of an energy-momentum tensor that encodes the corresponding quantum corrections. Therefore, we consider (\ref{eq:metric}) as a solution of the Einstein equations in the presence of an  ideal anisotropic fluid.
\begin{equation}
    T_{\mu\nu}=(\rho+p_t)u_{\mu}u_{\nu}+g_{\mu\nu}p_t+(p_r-p_t)s_{\mu}s_{\nu},
\end{equation}
where  $\rho$, $p_r$, $p_t$ are the fluid density, radial, and tangential pressure, respectively. The fluid velocity $u_{\mu}$, and radial space-like unit vector $s_{\mu}$ satisfy
\eq{
u_{\mu}u^{\mu}=-1,\,s_{\mu}s^{\mu}=1,\,u_{\mu}s^{\mu}=0.
}

The of the axial gravitational perturbations $h_{\mu\nu}$ in the Regge-Wheeler gauge \cite{Regge:1957td} takes the following form
%
%\begin{widetext}
\begin{eqnarray}
h^{axial}_{\mu \nu}= \left[
 \begin{array}{cccc}
 0 & 0 &0 & h_0
\\ 0 & 0 &0 & h_1
\\ 0 & 0 &0 & 0
\\ h_0 & h_1&0 &0
\end{array}\right]
\left(\sin\theta\frac{\partial}{\partial\theta}\right)
P_{\ell}(\cos\theta)\,, \label{pert_axial}
\end{eqnarray}
where $h_0(t,x)$ and $h_1(t,x)$ are two unknown functions, and $P_{\ell}(\cos\theta)$  is the Legendre polynomial. 
%\end{widetext}\
\subsection{Perturbations of test scalar and electromagnetic fields}

The equations governing the evolution of a scalar field $\Phi$ and the electromagnetic potential $A_\mu$ within the framework of general relativity can be expressed in the following form:
\begin{subequations}\label{coveqs}
\begin{eqnarray}\label{KGg}
\frac{1}{\sqrt{-g}} \partial_\mu \left( \sqrt{-g} g^{\mu \nu} \partial_\nu \Phi \right) &=& 0, 
\\\label{EmagEq}
\frac{1}{\sqrt{-g}} \partial_{\mu} \left( F_{\rho\sigma} g^{\rho \nu} g^{\sigma \mu} \sqrt{-g} \right) &=& 0\,,
\end{eqnarray}
\end{subequations}
where $F_{\mu\nu} = \partial_\mu A_\nu - \partial_\nu A_\mu$ represents the field strength tensor associated with the electromagnetic field.
By performing a separation of variables within the background metric (\ref{eq:metric_quasiglobal}), the equations of perturbations in all three cases can be transformed into the form of one scalar master wave-like equation with specific potential $V(x)$ \cite{Kokkotas:1999bd,Berti:2009kk,Konoplya:2011qq, Chen:2019iuo}:
\eq{
\left(\frac{\partial^2}{\partial t^2}-\frac{\partial^2}{\partial r_*^2}\right)\Psi+V_{i}(r_*)\Psi=0,\quad i=(s,v,ax),}
where $dx/dr_*=\sqrt{A/B}$ and
\eq
{V_{s}(x)=\frac{1}{2}\left(\frac{A}{B}\right)'\frac{R'}{R}+\frac{A}{B}\frac{R''}{R}+\frac{A}{R^2}\ell(\ell+1),
}
\eq{
V_{v}(x)=\frac{A}{R^2},
}
\begin{eqnarray}
V_{ax}(x) &=& \frac{A}{2B} \left[ 3\left(\frac{R'}{R}\right)^2 - \frac{R''}{R} \right] 
- \left(\frac{A}{2B} \frac{R'}{R} \right)' \nonumber \\
&& + A\frac{(\ell-1)(\ell+2)}{R^2},
\end{eqnarray}
for the scalar, vector fields and axial gravitational perturbations, correspondingly.

Using the following ansatz $$\Psi(t,r_*)=e^{-i\omega t}\Psi(r_*),$$
we obtain the master wave equation in the frequency domain
\eq{\label{eq:master_eq_freq}
\frac{d^2}{dr_*^2}\Psi+[\omega^2-V(r_*)]\Psi=0.
}

\begin{figure*}
\resizebox{\linewidth}{!}{\includegraphics{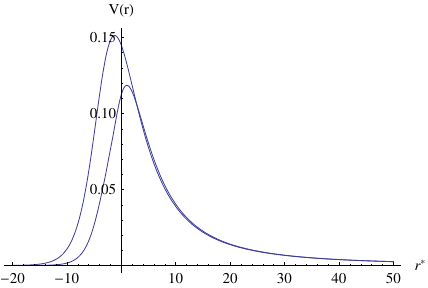}~~~\includegraphics{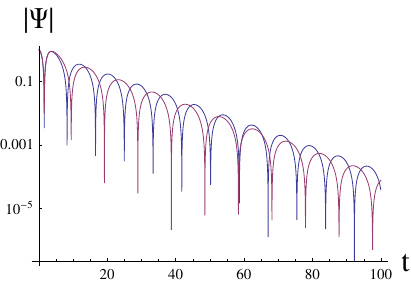}}
\caption{The effective potentials (left) and time-domain profiles for $\xi=0.1$ (upper, blue) and $\xi =3.9$ (lower, red), $\ell=2$, $M=1$.}\label{fig:TD2}
\end{figure*}

\begin{figure*}
\resizebox{\linewidth}{!}{\includegraphics{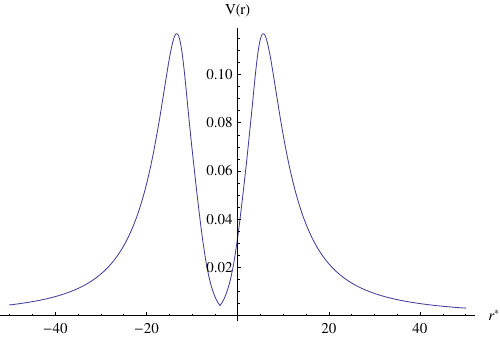}~~~\includegraphics{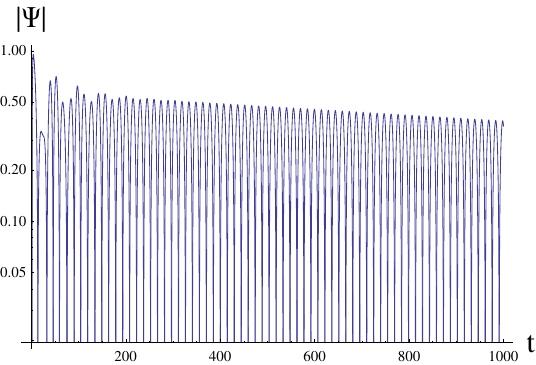}}
\caption{The effective potential (left) and the time-domain profile of perturbations (right) for $\xi=4$ (wormhole), $\ell=2$, $M=1$.}\label{fig:TD1}
\end{figure*}

The effective potential for axial gravitational perturbations is shown in figs. \ref{fig:TD2} and \ref{fig:TD1}. There one can see that the effective potentials are positive definite, which guarantees the stability of the black hole model at least within the framework of the considered effective types of perturbations.  
The effective potentials for the test scalar and vector fields exhibit qualitatively similar behavior.

\begin{figure*}
\resizebox{\linewidth}{!}{\includegraphics{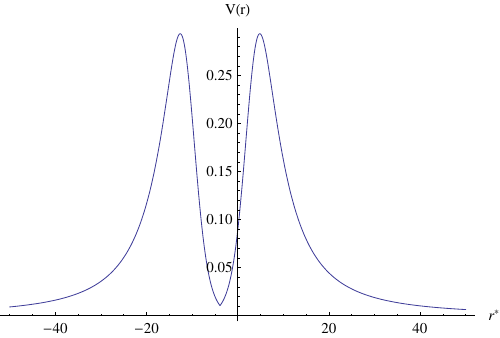}~~~\includegraphics{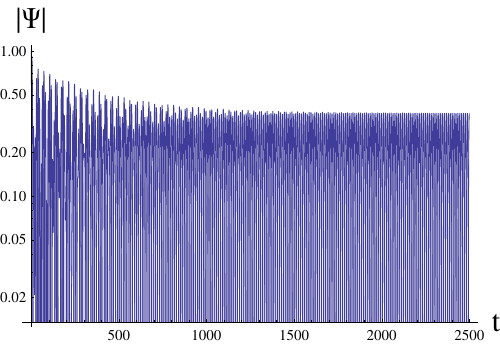}~~~\includegraphics{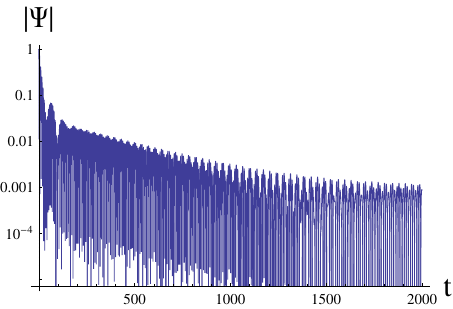}}
\caption{The effective potential (left) and the time-domain profiles of perturbations (middle and right) for $\xi=4$ (wormhole), $\ell=3$, $M=1$. The fundamental mode is $\omega_{0} = 0.288562 - 6.834\cdot10^{-6} i$ and the first overtone is $\omega_{1}=0.466434 - 0.002684 i$. For the middle profile, the observer is positioned between the two peaks, whereas for the right profile, the observer is located to the right of the peaks.}\label{fig:TD3}
\end{figure*}

\section{Methods}

Quasinormal frequencies $\omega$ are proper oscillation frequencies satisfying the boundary conditions,
\begin{equation}\label{boundaryconditions}
\Psi(r_*\to\pm\infty)\propto e^{\pm\imo \omega r_*},
\end{equation}
which are similar in the black hole and wormhole cases \cite{Konoplya:2005et} and correspond to purely incoming waves at the event horizon (BH) or at spatial infinity in the second universe (WH)  ($r_*\to-\infty$), and purely outgoing wave at spatial infinity ($r_*\to\infty$) in both cases, respectively.

Due to the symmetry of  $V_{i}$, which is $V_{i}(r_*)=V_{i}(-r_*)$, the corresponding solution $\Psi$ will be symmetric or antisymmetric with respect to the WH throat at $r_*=0$. Hence, we can reduce our boundary value problem for partial waves with fixed $\ell$ and $n$ in the WH case to the semi-interval $r_*\in[0,\infty)$ with $\Psi_{\ell n}(0)=0$ for the anti-symmetric case and $\frac{d}{dr_*}\Psi_{\ell n}|_{r_*=0}=0$ for the symmetric case. 
Notice that this does not mean that the full wave function must vanish on the throat and the perturbations must evolve independently on both sides of the throat. Indeed, when using time domain integration we do not impose vanishing of the perturbation at the throat and still we obtain the same quasinormal modes as via the pseudospectral method with vanishing partial wave of a given mode at the throat.

Here we will review the two methods used for finding quasinormal modes in time and frequency domians: pseudospectral method and time-domain integration.

\subsection{Pseudospectral method} 
To apply the Chebyshev pseudospectral method, we need to guarantee the correct behavior of the wave-function at the corresponding boundaries, that is, at the horizon and spatial infinity for the BH case and at spatial infinities at both universes for the WH case, respectively.
First of all, we  compactify our intervals $r\in [r_h,\infty)$ (BH) and $x\in(0,\infty)$ (WH) to the interval $(0,1)$ by introducing new variables
\eq{
r=\frac{r_h}{u},\quad\hbox{and}\quad x=\frac{u}{1-u},~~u\in(0,1),
}
Then we can separate singular contributions at the boundaries in the BH case
\eq{
\Psi^{BH}(r)=(1-u)^{-\frac{i\omega}{f'(r_h)}}u^{-2i \omega}e^{i r_h\omega/u}y(r),
}
and in the WH case
\eq{
\Psi^{WH}(x)=z(u)(1-u)^{-2i \omega}e^{\frac{i \omega}{1-u}}y(x),
}
where $z(u)=\sqrt{u}$ for the anti-symmetric solution and $z(u)=1$ for the symmetric one, respectively.

Then, we discretize equation (\ref{eq:master_eq_freq}) on the Chebyshev-Lobatto grid with nodes at 
\eq{u_j=\frac{1}{2}\left(1-\cos\left[\frac{\pi j}{N}\right]\right),~~j=0,1...N.}
This leads to the discretized matrix equation
\eq{\label{eq:discr_eq}
\left(\tilde{M}_0+\tilde{M}_1
\omega+\tilde{M}_2\,\omega^2\right)\tilde{y}=0,
}
where $\tilde{y}$  is the vector of the unknown function's values, $\tilde{M}_i$   are the numerical matrices of discretized coefficients at the collocation grid points.

The QNM spectrum can then be found by directly solving the corresponding matrix equation (\ref{eq:discr_eq}). To avoid spurious eigenvalues, we perform the calculations on two grids of different sizes and select only the overlapping values \cite{boyd2013chebyshev}.
The typical number of points is $200-350$, which is sufficient to determine the fundamental mode and the first few overtones with good accuracy.

\subsection{Time-domain integration}

By performing integration over time at a fixed radial coordinate, the temporal evolution of the wave function can be directly observed. For numerical implementation, we employ the discretization scheme introduced by Gundlach, Price, and Pullin \cite{Gundlach:1993tp}, which is formulated as:
\begin{eqnarray}
\Psi\left(N\right)&=&\Psi\left(W\right)+\Psi\left(E\right)-\Psi\left(S\right)\nonumber\\
&&- \Delta^2 V\left(S\right)\frac{\Psi\left(W\right)+\Psi\left(E\right)}{8} + \mathcal{O}\left(\Delta^4\right),\label{Discretization}
\end{eqnarray}
Here, the grid points are specified as follows: 
\begin{align*}
N &\equiv \left(u + \Delta, v + \Delta\right), \\
W &\equiv \left(u + \Delta, v\right), \\
E &\equiv \left(u, v + \Delta\right), \\
S &\equiv \left(u, v\right).
\end{align*}
These points correspond to nodes on a two-dimensional numerical grid where $N$, $W$, $E$, and $S$ indicate the north, west, east, and south locations, respectively. The parameter $\Delta$ denotes the finite step size in both null directions $u$ and $v$. The potential function at point $S$ is represented by $V(S)$. Higher-order corrections proportional to $\Delta^4$ ensure improved accuracy for sufficiently small step sizes.

To extract the quasinormal frequencies from the time-domain evolution, we utilize the Prony method. This method approximates the time-domain signal by fitting a sum of exponentially damped oscillatory modes:
\begin{equation}
\Psi(t) \approx \sum_{i=1}^{p} C_i e^{-i\omega_i t}.\label{damping-exponents}
\end{equation}
In this expression, $C_i$ represents the amplitude of each mode, while $\omega_i$ denotes the complex frequency, which contains both the oscillation frequency and the decay rate. The imaginary part of $\omega_i$ characterizes the damping, while the real part determines the oscillatory behavior. By selecting the onset of the dominant ringdown phase, the quasinormal frequencies can be accurately retrieved from the waveform. This analysis is essential for understanding the response of black holes and compact objects to perturbations, offering insights into their stability and the nature of the surrounding spacetime.

\section{Quasinormal modes}

Here, we consider the properties of quasinormal modes spectra of black holes and wormholes, which, despite sharing the same boundary conditions, exhibit distinct characteristics. This difference arises because, in the wormhole case, the effective potential features a double peak, leading to modifications in the signal due to multiple reflections between the two peaks.  The study of pertrubations and QNM spectra for various wormhole models under different types of perturbations has been done in numerous papers (see, for instance, \cite{Khoo:2024yeh,Alfaro:2024tdr,Konoplya:2005et,Batic_2024,Azad:2022qqn,Churilova:2019cyt,Churilova:2019qph,Churilova:2021tgn,Konoplya:2018ala,Aneesh:2018hlp,Cuyubamba:2018jdl,Blazquez-Salcedo:2018ipc,Jusufi:2020mmy,Gonzalez:2022ote,Konoplya:2025mvj,Roy:2021jjg,Volkel:2018hwb,Bueno:2017hyj,Konoplya:2010kv,Bronnikov:2012ch,Bronnikov:2021liv}), but it is mostly restricted to the fundamental modes. Here, we extend the analysis to investigate the behavior of higher overtones as well.

In this paper, we consider effective axial gravitational perturbations in greater detail than perturbations of test fields, as they are the ones that can potentially be detected by LIGO/LISA missions \cite{Arun_2022}.

For numerical calculations here and in the following sections, it is convenient to set \( M = 1 \). 1. The accurate values
of the QNMs are provided in Appendix \ref{sec:appendix}.

\subsection{Black hole}
\begin{figure*}
\resizebox{0.9 \linewidth}{!}{\includegraphics{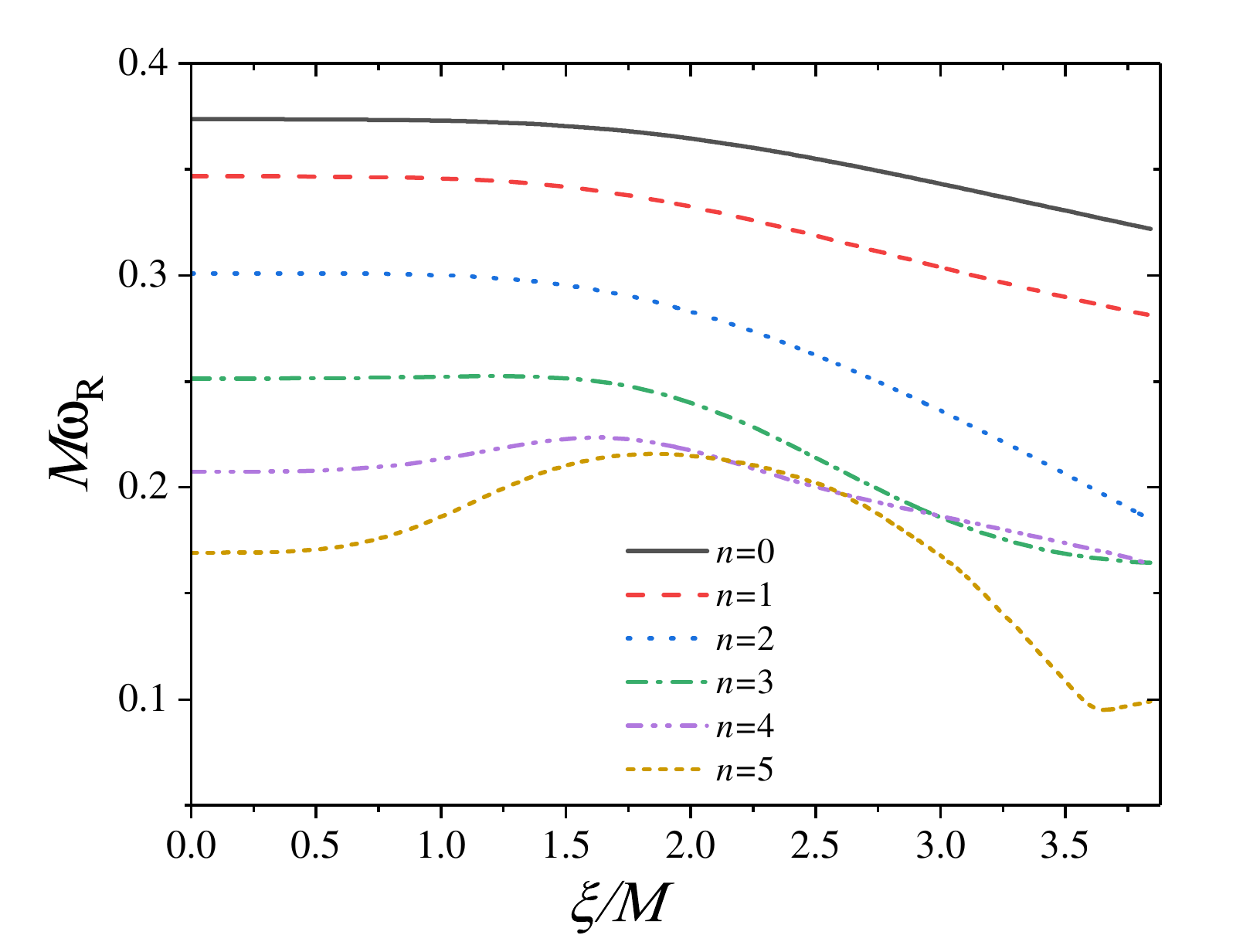}~~~\includegraphics{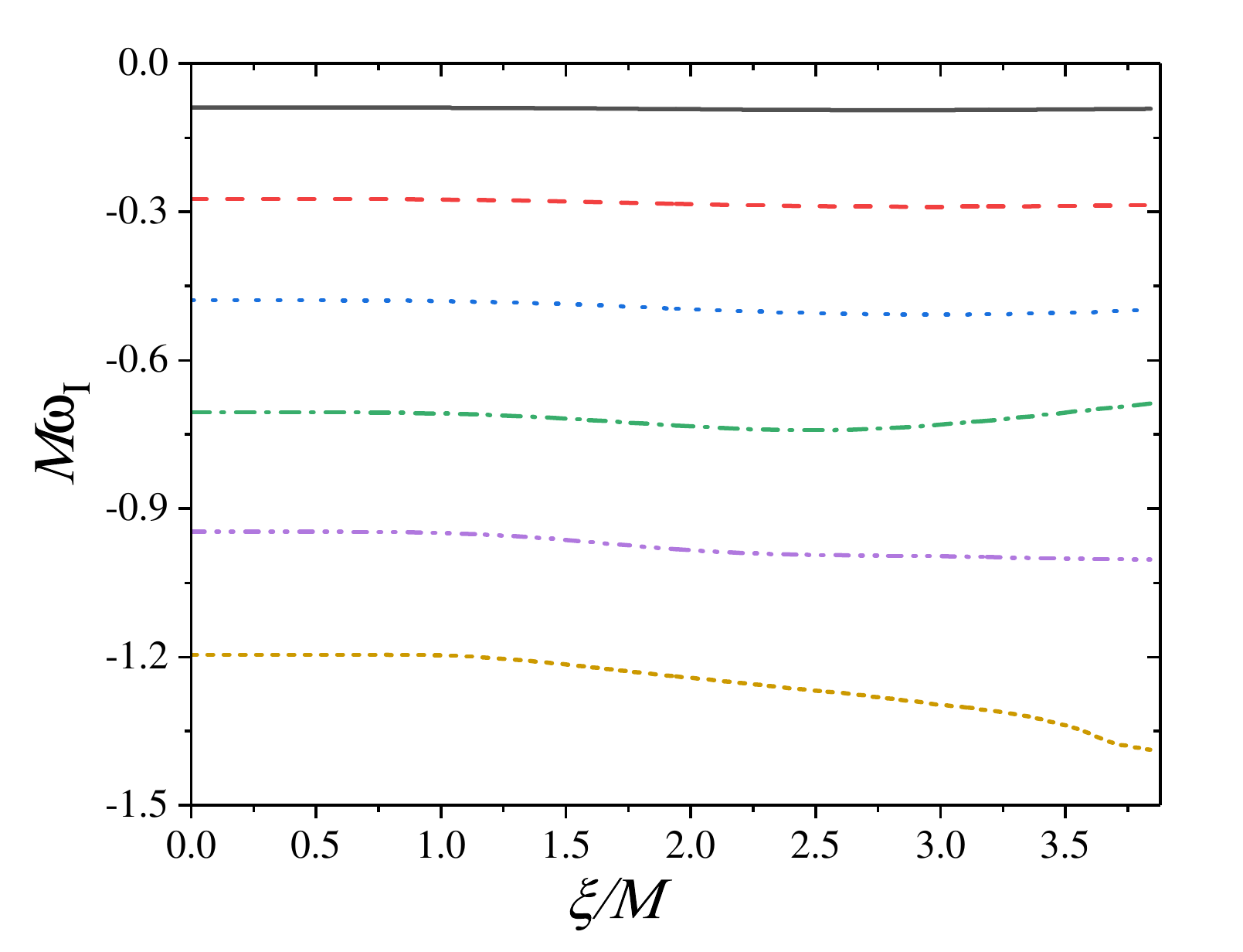}}
\caption{The fundamental mode and the first five overtones as a function of $\xi$  for $\ell=2$.}\label{fig:QNMs_axial}
\end{figure*}

\begin{figure}
\resizebox{\linewidth}{!}{\includegraphics{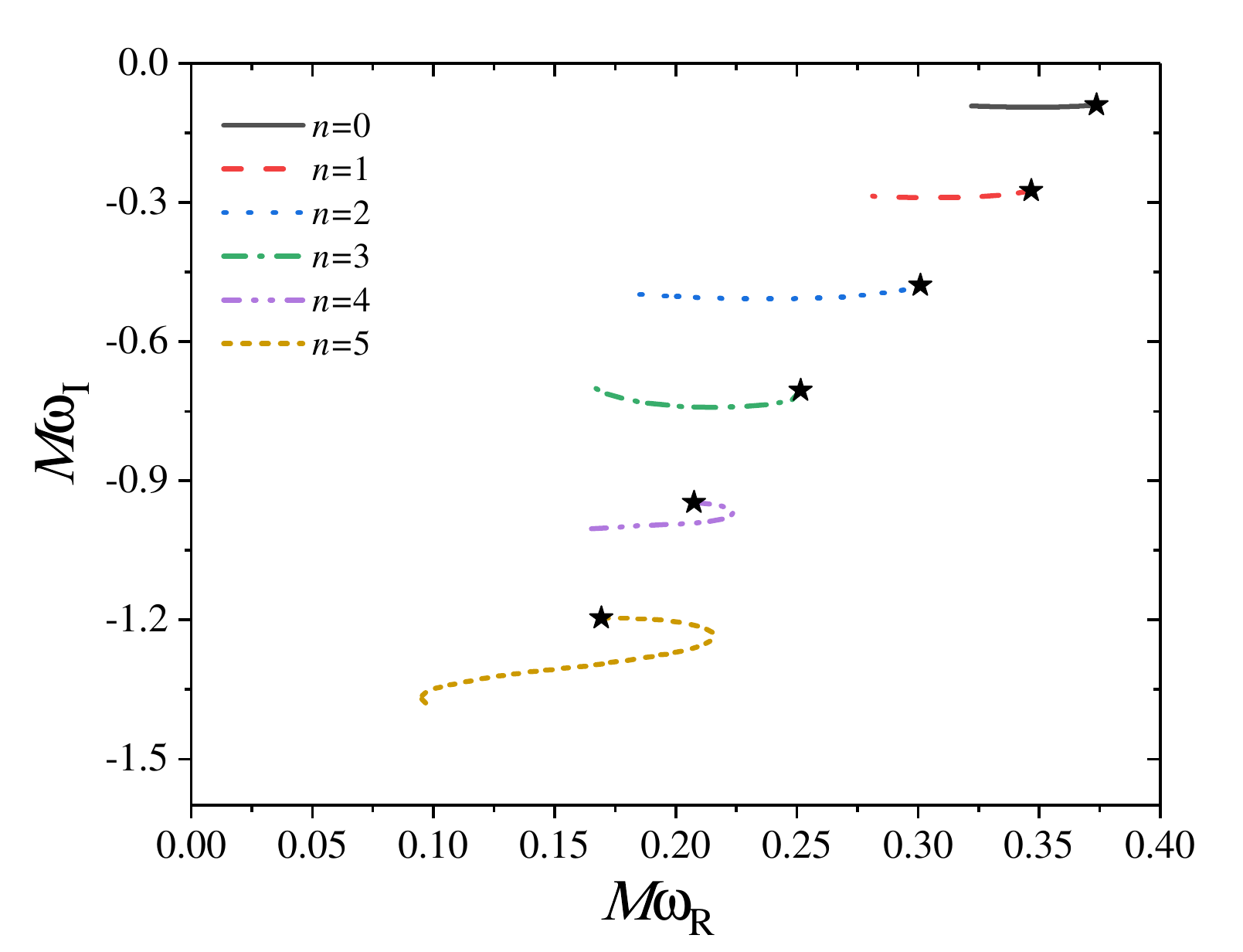}}
\caption{The QNMs trajectories in the complex $\omega$ plane for
the fundamental QNMs and the first five overtones, $\ell =2$. The black stars represent the Schwarzschild black hole  QNM values.}\label{fig:QNMsBH_OmegaPlane}
\end{figure}

\begin{figure*}
\resizebox{0.9 \linewidth}{!}{\includegraphics{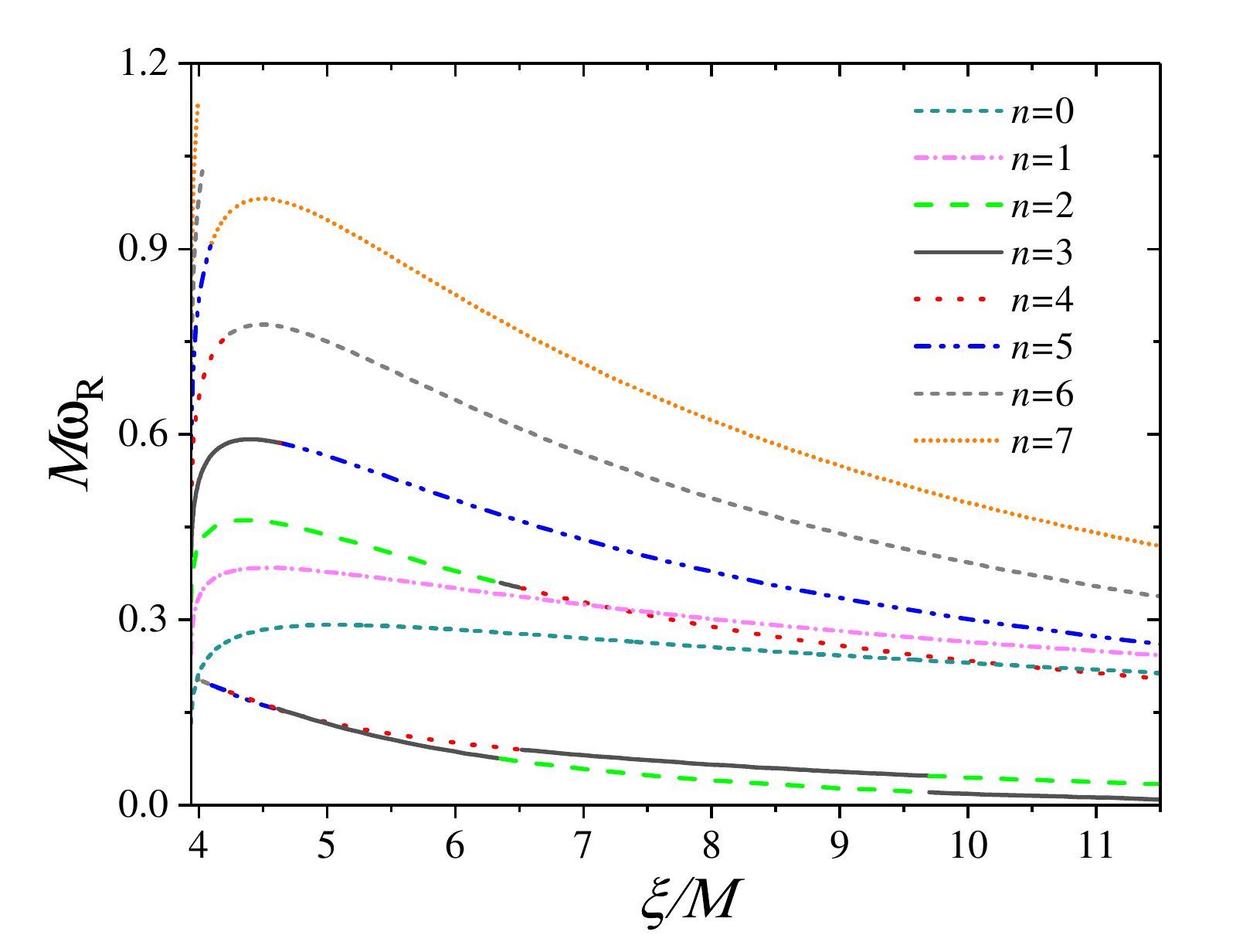}~~~\includegraphics{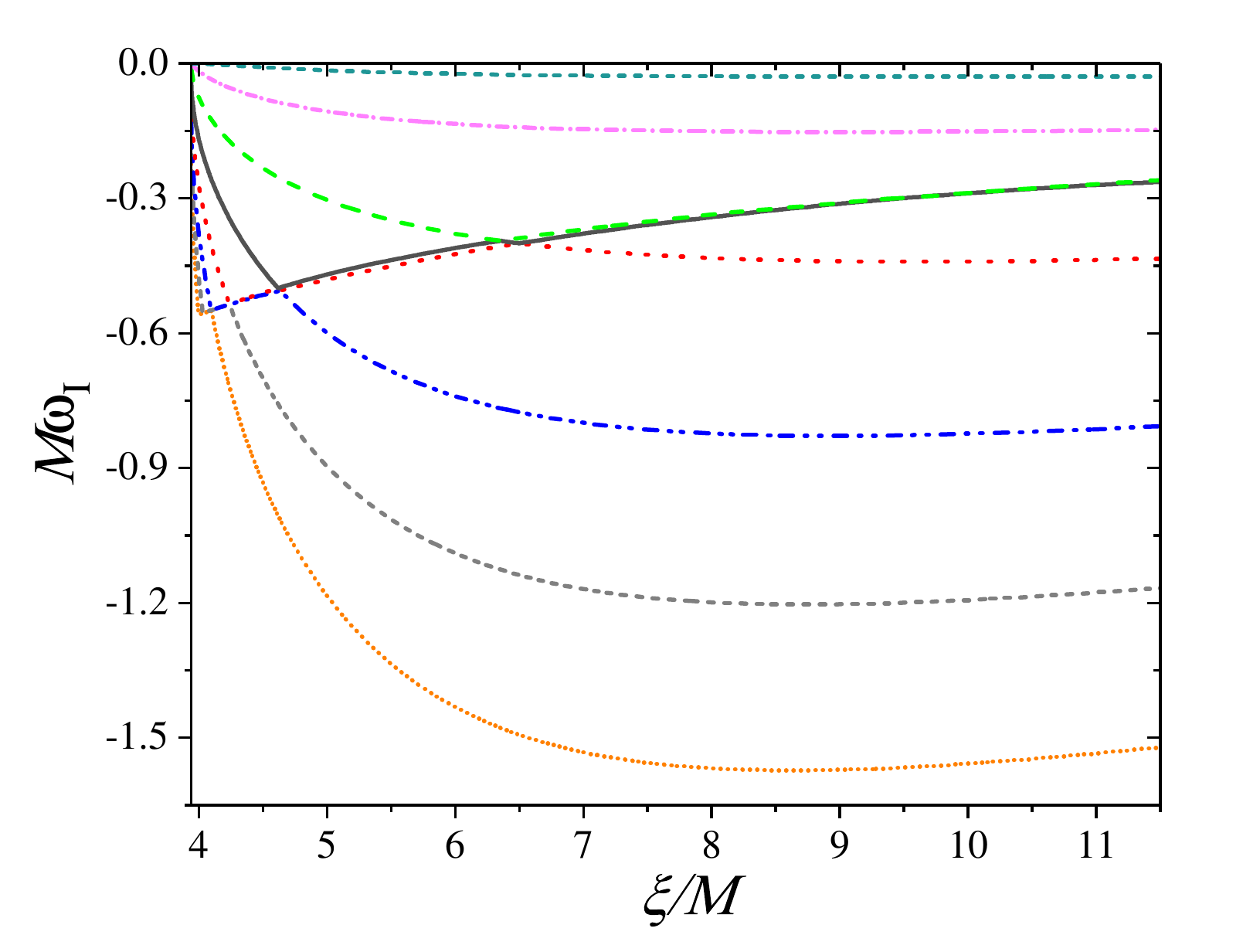}}
\caption{The fundamental mode and the first seven overtones as a function of $\xi$  for $\ell=2$ in the wormhole case.}\label{fig:QNMs_axialWH}
\end{figure*} 
\begin{figure}
\resizebox{0.9 \linewidth}{!}{\includegraphics{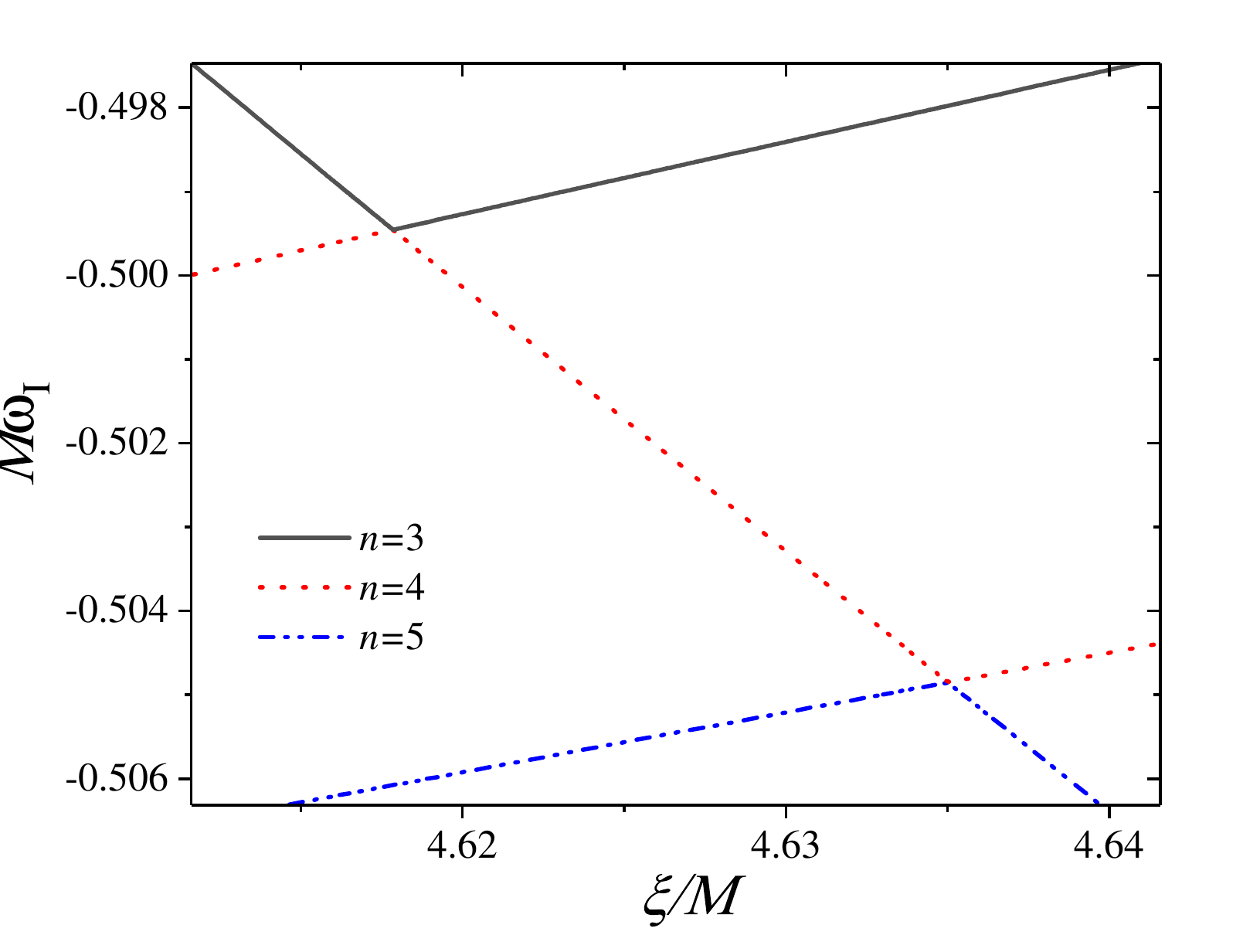}~~~}
\caption{The typical behavior of overtone curves near the reconnection points, $\ell=2$.}\label{fig:qnmsWH_recconnection}
\end{figure}
\begin{figure}
\resizebox{\linewidth}{!}{\includegraphics{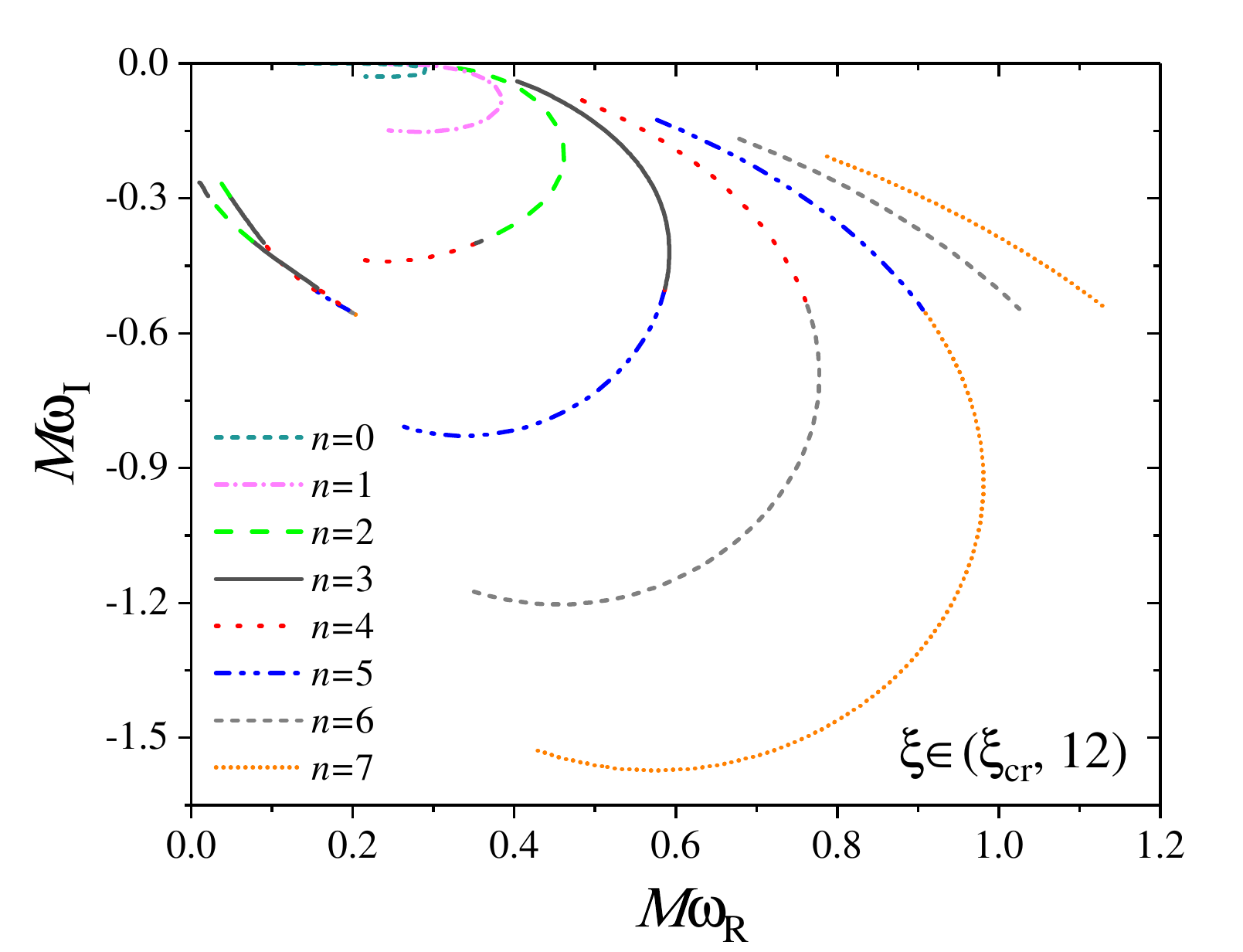}}
\caption{The QNMs trajectories in the complex $\omega$ plane for
the fundamental QNMs and the first seven overtones, $\ell =2$.}\label{fig:QNMsWH_omega}
\end{figure}

The precise values of the fundamental mode and the first few overtones of the QNM spectra for effective axial gravitational perturbations, obtained using the pseudospectral method, are presented in Table \ref{tab:QNMs_BH} for \( \ell = 2 \) and \( \ell = 3 \). In the first part of Tables \ref{tab:QNMs_WHScalar}–\ref{tab:QNMs_AxBHWHtd}, we provide the fundamental mode values obtained by fitting waveform profiles from time-domain integration. Typical examples of the waveform profiles are shown in Fig. \ref{fig:TD2} (right panel).  The dependence of the QNM spectra for effective axial perturbations on \( \xi \) is illustrated in Fig. \ref{fig:QNMs_axial}, while the corresponding trajectories in the complex \( \omega \) plane are shown in Fig. \ref{fig:QNMsBH_OmegaPlane}. It can be observed that the real part of \( \omega \) for the \( n = 0,1,2 \) modes decreases monotonically up to \( \xi = \xi_{cr} \), whereas modes with \( n > 2 \) initially increase monotonically up to a certain \( \xi \) before exhibiting nonmonotonic behavior, undergoing rapid and significant variations compared to the \( \xi = 0 \) case. 
The imaginary part of the corresponding modes also follows a nonmonotonic trend; however, in contrast to the real part, its deviation from the \( \xi = 0 \) case occurs at a much slower rate.

From the data on the first few overtones obtained using the pseudospectral method for the black hole configuration, we observe that while the fundamental mode ($n = 0$) changes relatively mildly—by about $10-15\%$—the third overtone deviates from its Schwarzschild limit by more than $30\%$, with the deviation increasing as the overtone number grows. For smaller multipole numbers $\ell$, which correspond to the dynamical degrees of freedom of fields with lower spin (such as scalar, electromagnetic, and Dirac fields), the deviation of overtones from their Schwarzschild limits is expected to be even stronger, as noted in \cite{Konoplya:2024lch}.

\subsection{Wormhole}
The results for the wormhole case, as expected, differ significantly from those for the black hole case. Similar to the BH case, we provide the precise values of the fundamental mode and the first few overtones of the QNM spectra for effective axial gravitational perturbations in Table \ref{tab:QNMs_WH} for \( \ell = 2 \) and \( \ell = 3 \). In the second part of Tables \ref{tab:QNMs_WHScalar}–\ref{tab:QNMs_AxBHWHtd}, we present the corresponding values of the fundamental mode obtained from time-domain integration. Fundamental modes for test scalar and electromagnetic fields are provided in Tables \ref{tab:QNMs_WHScalar} and \ref{tab:QNMs_WHElmag}. Typical examples of waveform profiles are shown in Figs. \ref{fig:TD1}–\ref{fig:TD3}.

For values of \( \xi \) near the BH/WH transition, an intriguing phenomenon arises, characterized by the presence of arbitrarily long-lived fundamental modes and a sequence of echoes during the early stage of the ringdown phase. These echoes are followed by a regular ringdown phase at late times, dominated by a slowly decaying quasinormal frequency. However, the time-domain evolution of the early-stage perturbation changes smoothly across the transition. For larger \( \ell \), the duration of the echo phase is extended, and the onset of the regular ringdown phase is delayed (see Fig. \ref{fig:TD3} for the \( \ell = 3 \) case). Nevertheless, at the final stage of the ringdown, the long-lived modes become dominant for larger \( \ell \) as well.
As a result, the signal modification primarily occurs at late times, while the early ringdown phase remains similar to that of a black hole near the transition. The presence of such modes is not unique to wormholes alone \cite{Cardoso:2016rao,Bronnikov:2019sbx,Churilova:2019cyt} but is also observed in other types of horizonless exotic compact objects \cite{Stashko:2023ffs,Chirenti:2012fr,Franzin_2024}. Similar modes can also appear in the presence of massive field perturbations \cite{Ohashi:2004wr,Konoplya:2024wds,Konoplya:2004wg,Zinhailo:2024jzt,Davlataliev:2024mjl}.

However, there is a crucial distinction between these cases: the arbitrarily long-lived modes of massive fields do not manifest in the time-domain profile of the perturbation due to the early onset of asymptotic oscillatory tails with a power-law envelope. In contrast, the arbitrarily long-lived modes of wormholes at the threshold of their formation are clearly visible in the evolution of perturbations.

While these oscillatory tails are of interest—particularly because they contribute to very low-frequency gravitational waves \cite{Konoplya:2023fmh} detected by the Pulsar Timing Array experiment \cite{Konoplya:2023fmh}—they suppress quasinormal modes of massive fields, including arbitrarily long-lived quasinormal modes, rendering the latter practically unobservable. Therefore, if very long-lived quasinormal modes are detected in the future, this may indicate the transition of a black hole into a wormhole state.

The second intriguing feature of the wormhole is the behavior of the QNM spectrum. The modes do not exhibit a smooth transition and instead form a completely different set of modes, including long-lived modes described below. This indicates that the frequencies are non-perturbative in $\xi$, as they do not transition into the Schwarzschild-like modes of the quantum-corrected black hole for smaller values of $\xi$.
The typical dependencies of the real and imaginary parts of the quasinormal modes as function of $\xi$ are shown in Fig. \ref{fig:QNMs_axialWH} and the corresponding curves in the complex $\omega$ plane are shown in Fig. \ref{fig:QNMsWH_omega}.  While the real part ${\rm Re}(\omega_n)$ and the imaginary part $|{\rm Im}(\omega )|$ of the $n=0$ and $n=1$ branches are bounded from above and monotonically decrease with increasing $\xi$, a much more complex behavior emerges in the higher overtone branches with $n > 2$. At specific point, lets say $\xi = \xi_i$, the imaginary parts of two successive frequencies, $\omega_n$ and $\omega_{n+1}$, coincide such that ${\rm Im}(\omega_n) ={\rm Im}(\omega_{n+1})$. For $\xi > \xi_i$, we have ${\rm Im}(\omega_n) < {\rm Im}(\omega_{n+1})$ which leads to reassigning of their overtone numbers $n\leftrightarrow n+1$ (see Fig. \ref{fig:qnmsWH_recconnection}), leading to a jump in the real part of the frequency.
It is worth mentioning that similar effect of QNMs modes' reconnection  can be also observed in scenarios where QNM spectrum  undergoes abrupt changes due to relatively small static deformations of the background or the presence of additional peaks in the effective potential, which induce secondary scatterings \cite{Cheung_2022}.

In the eikonal limit and at several orders beyond it one can derive analytic approximate expressions for quasinormal modes using the general approach developed in \cite{Konoplya:2023moy} and consequently used in a number of works \cite{Malik:2024tuf,Malik:2024bmp,Malik:2024sxv,Malik:2024elk}. Using the series expansion in terms of powers of $1/(\ell+(1/2))$ \cite{Konoplya:2023moy}, we find the expansion for the position of a maximum of the effective potential for the scalar field,
\begin{equation}\label{rmax-scalar}
\begin{aligned}\nonumber
r_{\max } = & \ 3 M - \frac{M}{3 \kappa^2} \\
           & + \xi^4 \left(\frac{2}{243 M^3} - \frac{58}{6561 M^3 \kappa^2}\right) 
             + \mathcal{O}\left(\frac{1}{\kappa^4},\xi^6\right),
\end{aligned}
\end{equation} 
and, using the WKB formula developed in \cite{Konoplya:2003ii,Iyer:1986np,Schutz:1985km}, the expression for the quasinormal modes\\
\begin{widetext}
\begin{equation}\label{eikonal-scalar}
\begin{array}{rcl}
\omega  &=& \displaystyle-\frac{i K \left(441004 K^2-123815\right)}{46656 \sqrt{3} M \kappa ^2}-\frac{3516 K^2+43}{1296 \sqrt{3} M \kappa }+\frac{\kappa }{3 \sqrt{3} M}-\frac{i K}{3 \sqrt{3} M}\\
&&\displaystyle+\xi ^4 \left(\frac{i K \left(3902948 K^2-1348285\right)}{51018336 \sqrt{3} M^5 \kappa ^2}+\frac{4956 K^2+547}{472392 \sqrt{3} M^5 \kappa }-\frac{2 \kappa }{6561 \sqrt{3} M^5}-\frac{10 i K}{6561 \sqrt{3} M^5}\right)+\mathcal{O}\left(\frac{1}{\kappa ^3},\xi ^6\right).
\end{array}
\end{equation}
Here we used $\kappa\equiv\ell+1/2$, $K\equiv n+1/2$).

In a similar way, the position of the maximum for the electromagnetic field is
\begin{equation}\label{rmax-electromagnetic}
r_{\max } = 3 M+\frac{2 \xi ^4}{243 M^3}+\mathcal{O}\left(\frac{1}{\kappa ^4},\xi ^6\right),
\end{equation}
and the quasinormal modes are
\begin{equation}\label{eikonal-electromagnetic}
\begin{array}{rcl}
\omega  &=& \displaystyle-\frac{i K \left(441004 K^2-73703\right)}{46656 \sqrt{3} M \kappa ^2}-\frac{3516 K^2+187}{1296 \sqrt{3} M \kappa }+\frac{\kappa }{3 \sqrt{3} M}-\frac{i K}{3 \sqrt{3} M}\\
&&\displaystyle+\xi ^4 \left(\frac{i K \left(3902948 K^2-1268797\right)}{51018336 \sqrt{3} M^5 \kappa ^2}+\frac{7 \left(708 K^2+37\right)}{472392 \sqrt{3} M^5 \kappa }-\frac{2 \kappa }{6561 \sqrt{3} M^5}-\frac{10 i K}{6561 \sqrt{3} M^5}\right)+\mathcal{O}\left(\frac{1}{\kappa ^3},\xi ^6\right).
\end{array}
\end{equation}
\\
For axial gravitational perturbations we have:
\begin{equation}\label{rmax-scalar}
r_{\max } = \frac{M}{\kappa ^2}+3 M+\xi ^4 \left(\frac{26}{2187 M^3 \kappa ^2}+\frac{2}{243 M^3}\right)+\mathcal{O}\left(\frac{1}{\kappa ^4},\xi ^6\right),
\end{equation}
\begin{equation}\label{eikonal-scalar}
\begin{array}{rcl}
\omega  &=& \displaystyle-\frac{i K \left(441004 K^2-47783\right)}{46656 \sqrt{3} M \kappa ^2}-\frac{3516 K^2+619}{1296 \sqrt{3} M \kappa }+\frac{\kappa }{3 \sqrt{3} M}-\frac{i K}{3 \sqrt{3} M}\\
&&\displaystyle+\xi ^4 \left(\frac{i K \left(3902948 K^2-2419645\right)}{51018336 \sqrt{3} M^5 \kappa ^2}+\frac{4956 K^2-29}{472392 \sqrt{3} M^5 \kappa }-\frac{2 \kappa }{6561 \sqrt{3} M^5}-\frac{10 i K}{6561 \sqrt{3} M^5}\right)+\mathcal{O}\left(\frac{1}{\kappa ^3},\xi ^6\right)
\end{array}
\end{equation}
\end{widetext}

The analytic expressions above provide an approximation for quasinormal modes at relatively small values of \( \xi \) and sufficiently large values of \( \ell \). One can easily verify that, in the eikonal limit, these expressions uphold the correspondence between null geodesics and quasinormal modes \cite{Cardoso:2008bp}, despite the existence of several cases where this correspondence breaks down \cite{Konoplya:2020bxa,Konoplya:2017wot,Konoplya:2022gjp,Bolokhov:2023dxq}.

\begin{figure*}
\resizebox{0.9 \linewidth}{!}{\includegraphics{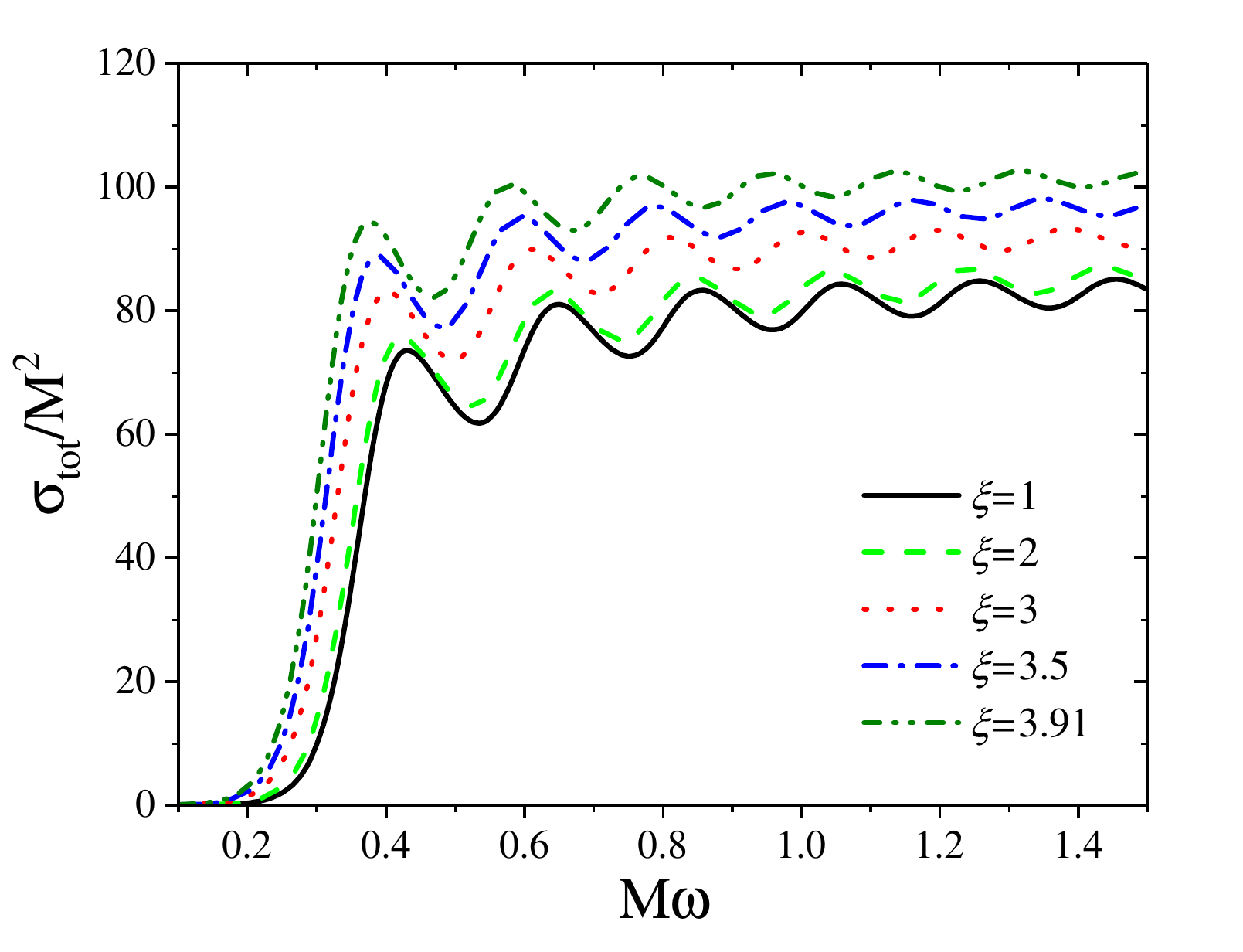}~~~\includegraphics{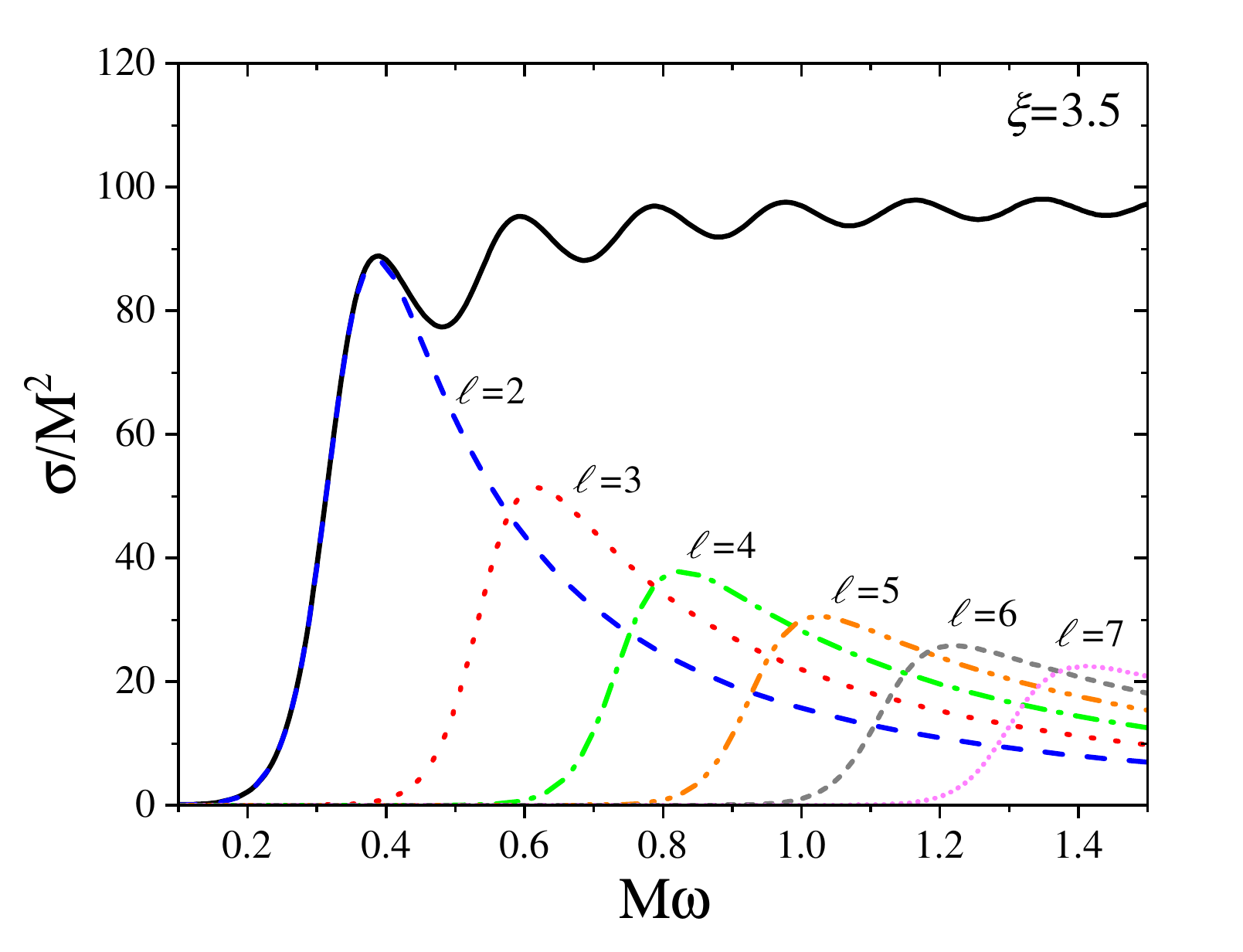}}
\caption{The typical behavior of the total absorption cross-section for different values of  $\xi$ in the BH case (left panel).  For the lower values of $\xi$, the corresponding curves are practically indistinguishable. The contributions of the partial cross-sections with different $\ell$ to the total cross-section for  $\xi=3.5$ (right panel).
}\label{fig:crossectionBH}
\end{figure*}
\begin{figure}
\resizebox{0.9 \linewidth}{!}{\includegraphics{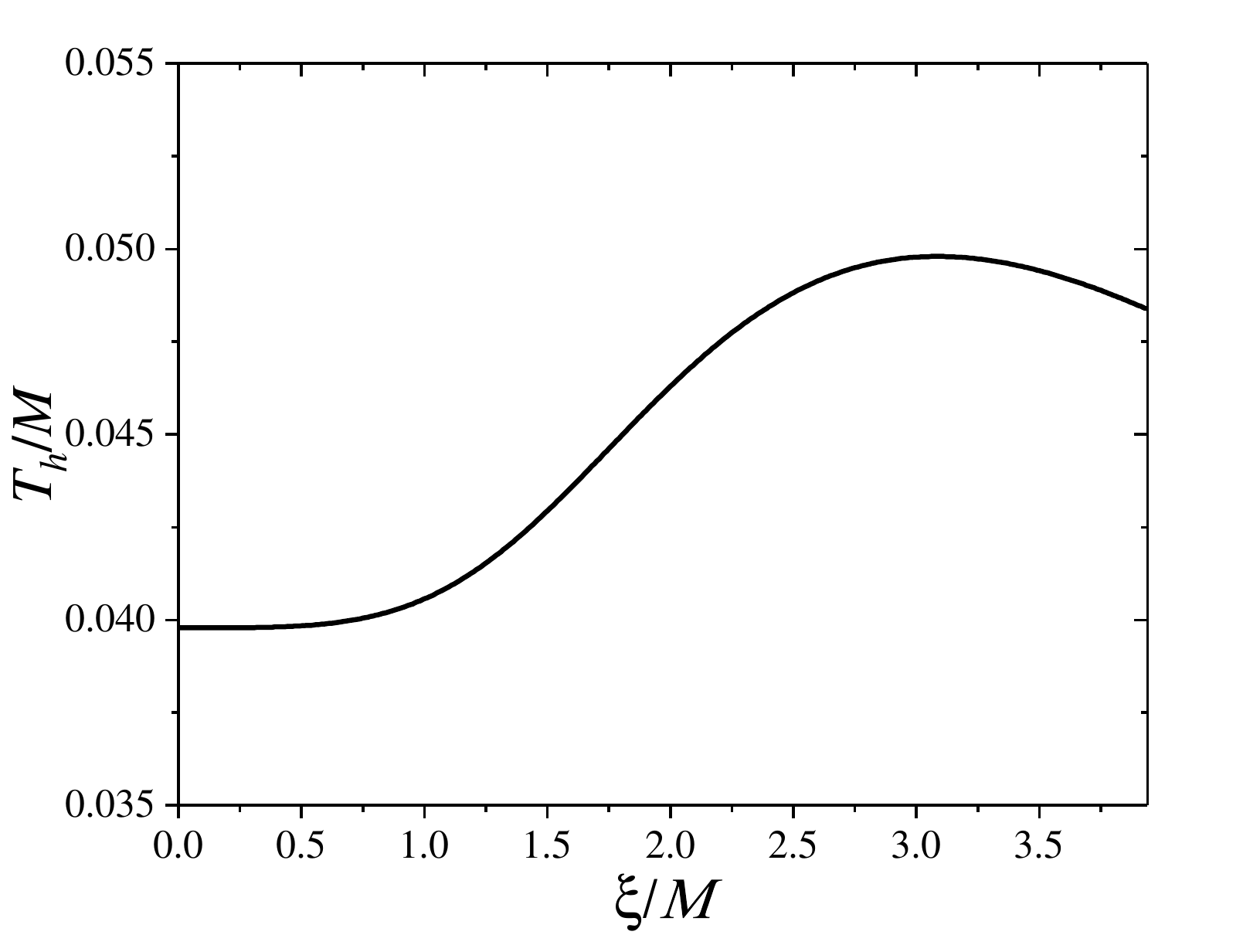}~~~}
\caption{The Hawking temperature as  a function of $\xi$. There is a maximum at $\xi\approx 3.08M$}\label{fig:Hawk_temp}
\end{figure}
\begin{figure*}
\resizebox{0.9 \linewidth}{!}{\includegraphics{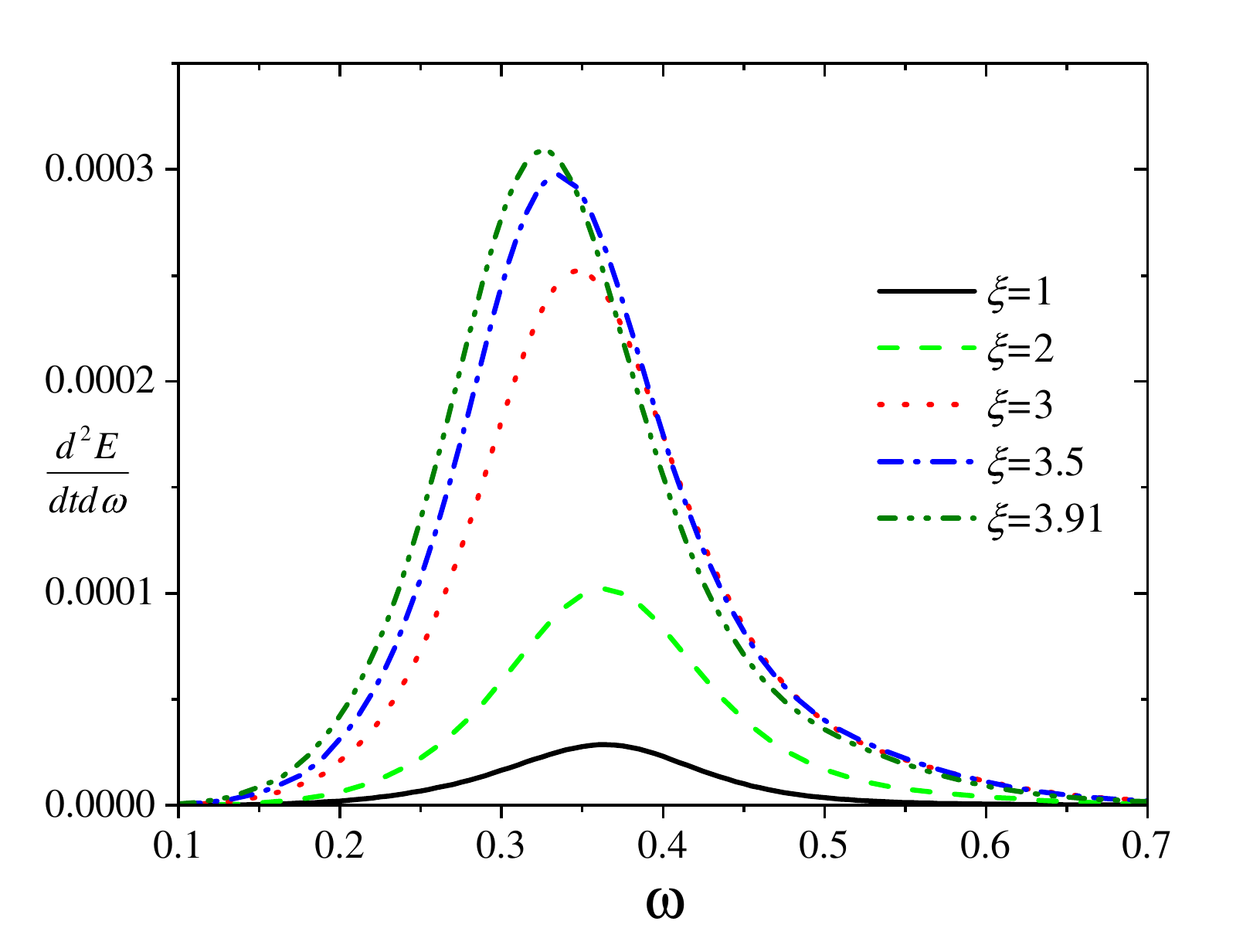}~~~\includegraphics{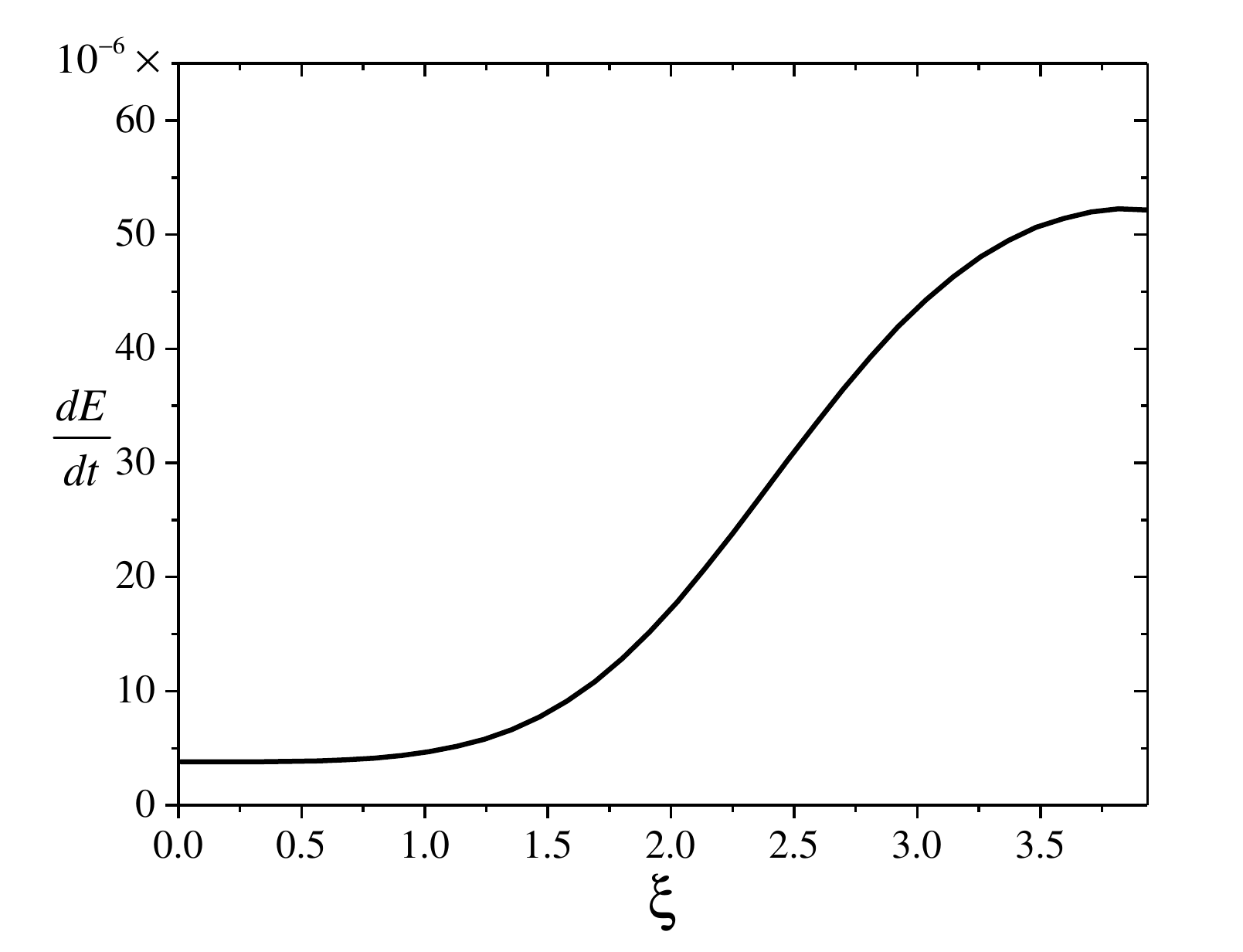}}
\caption{Typical dependencies of the energy emission rate per unit frequency for various values of $\xi$ as function of $\omega$ (left) and  the total energy emission rate (right). For the total energy emission rate there is a maximum at $\xi\approx3.86$.
}\label{fig:Hawking_radiation}
\end{figure*}

\begin{figure*}
\resizebox{0.9 \linewidth}{!}{\includegraphics{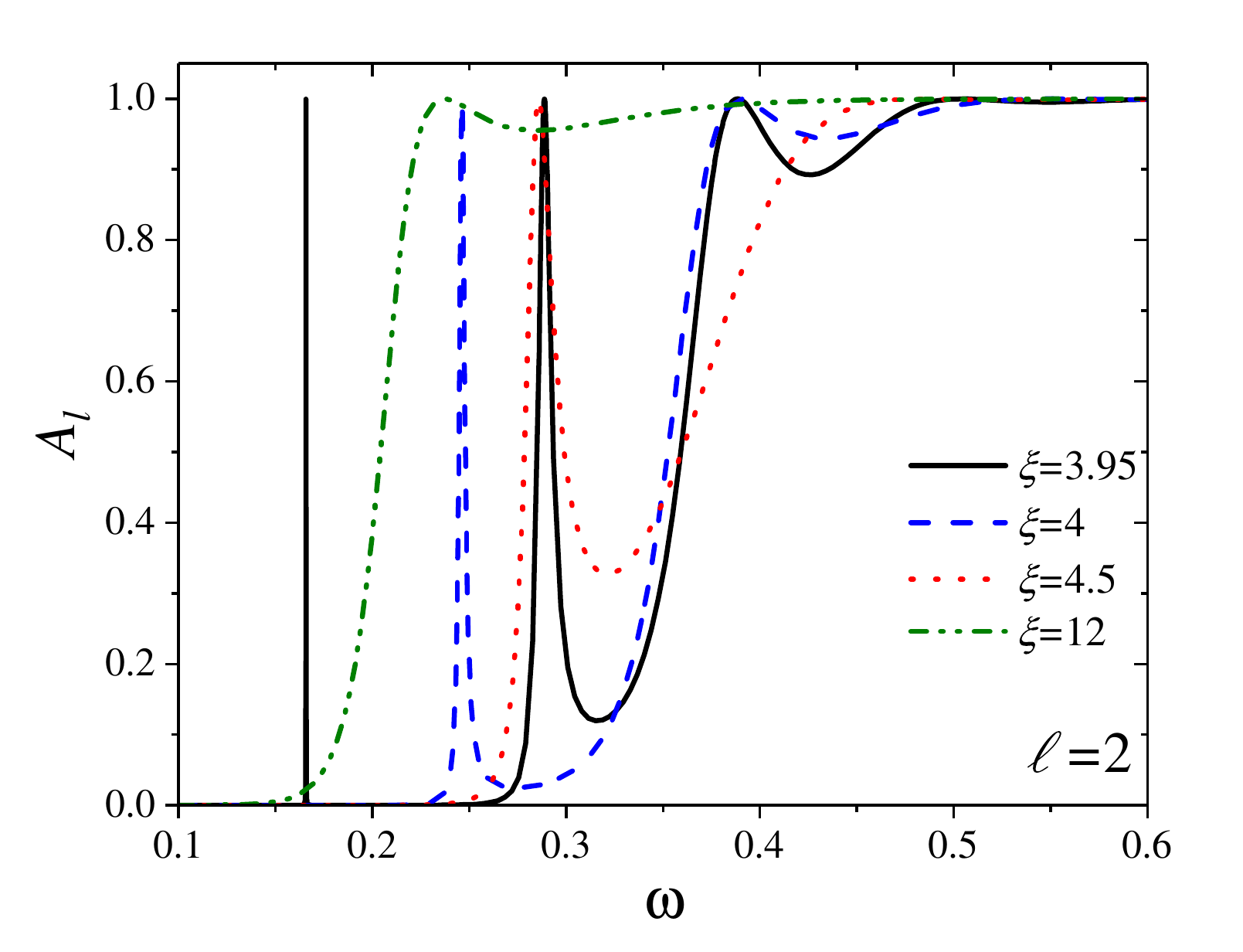}~~~\includegraphics{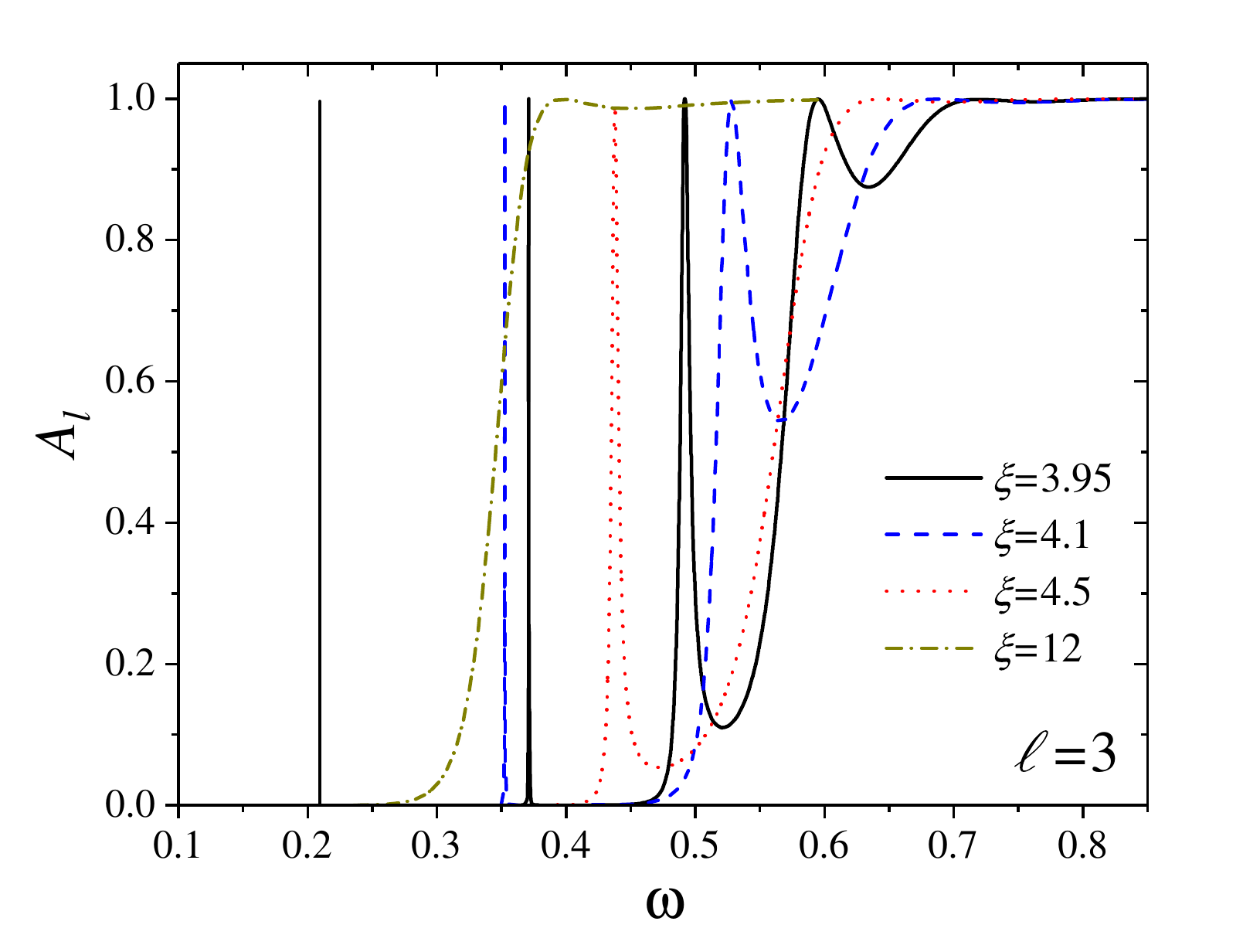}}
\caption{Typical behavior of  greybody factors for various values of  $\xi$ in the WH case: $\ell=2$ and $\ell=3$. For some specific values of $\omega$, one can observe presence of trapped modes, which are represented by the narrow vertical peaks.}\label{fig:GB_WH}
\end{figure*}

\begin{figure*}
\resizebox{0.9 \linewidth}{!}{\includegraphics{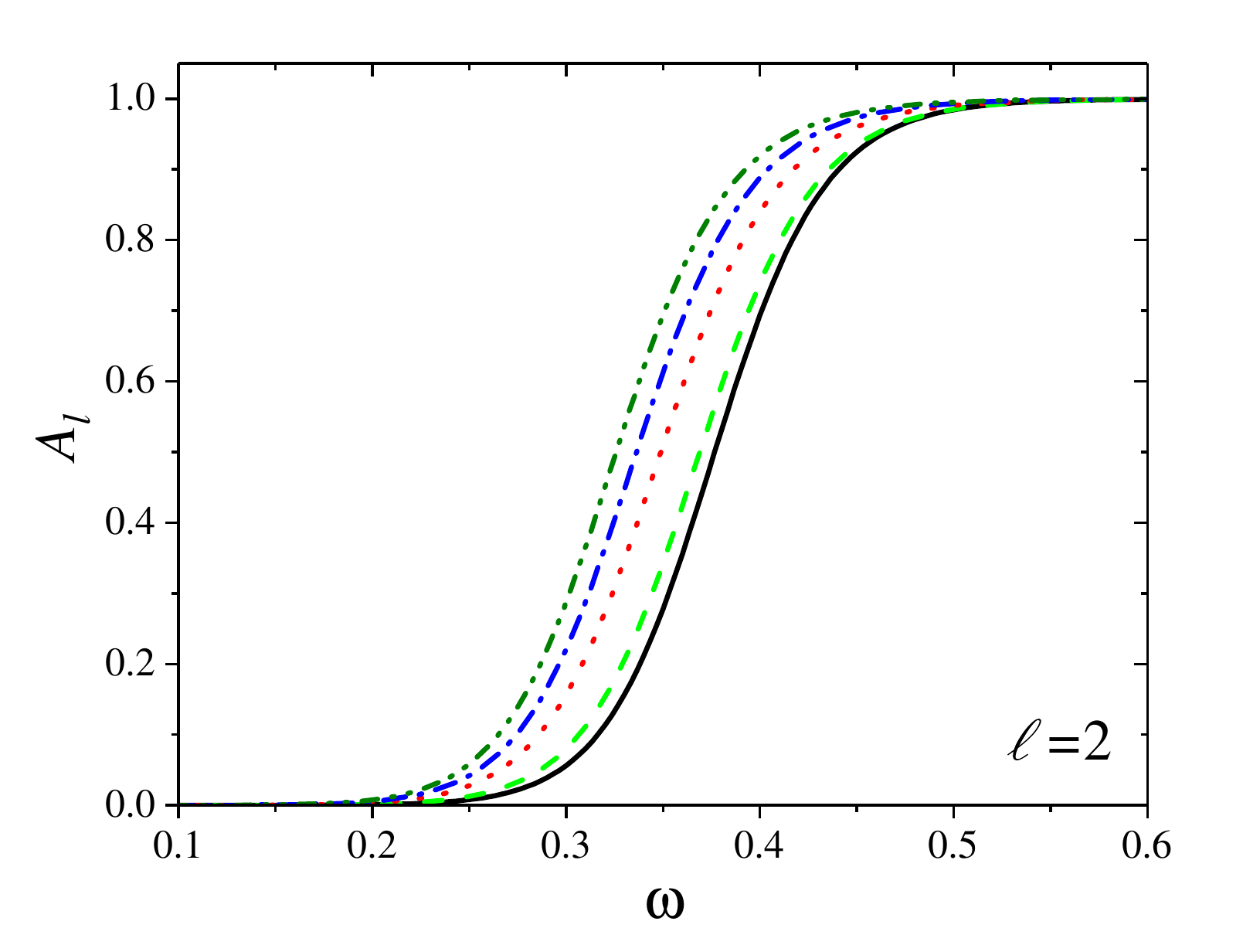}~~~\includegraphics{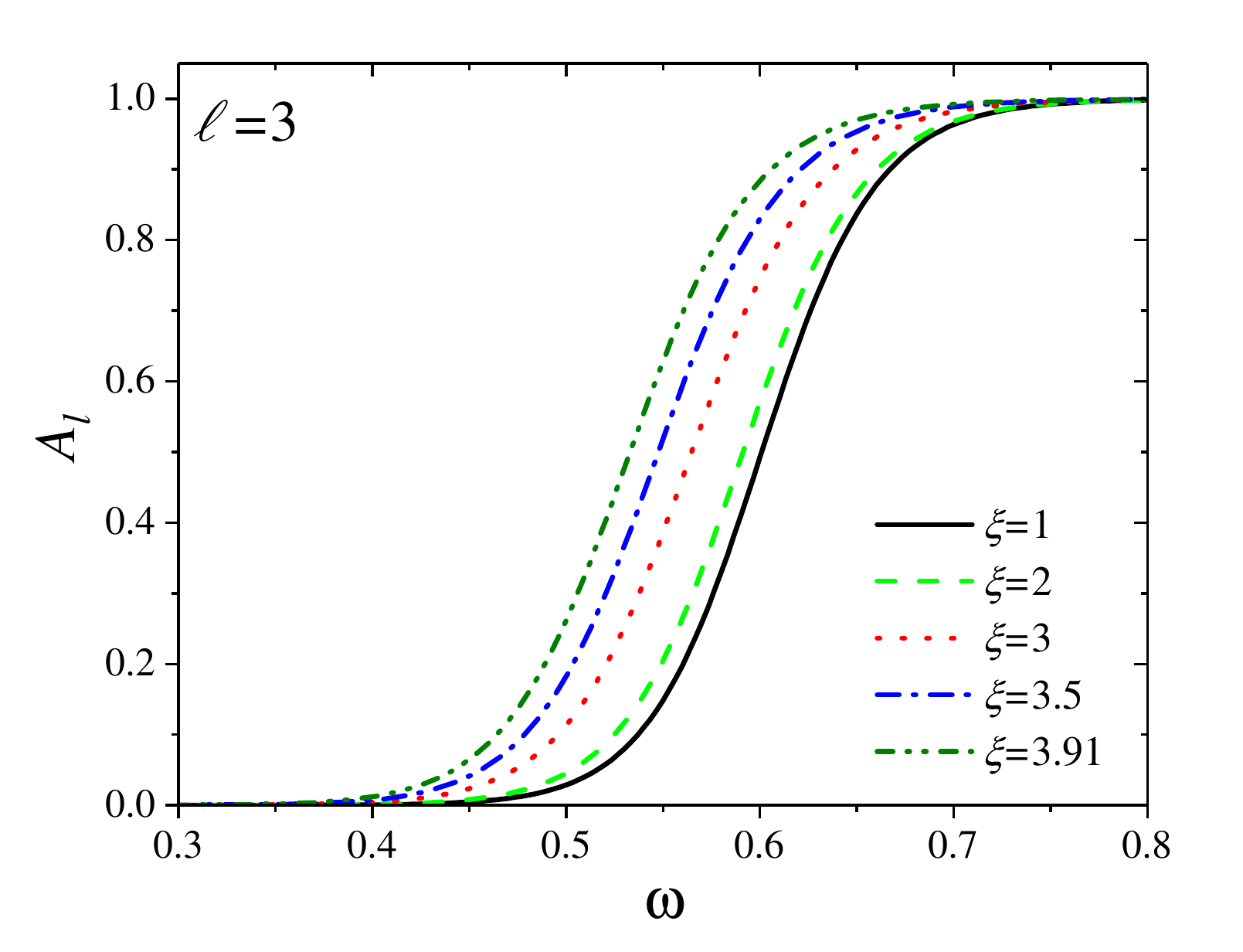}}
\caption{The typical behavior of the grey-body factors for different values of $\xi$ with $\ell=2$ (left panel) and $\ell=3$ (right panel) in the
black hole case. For the lower values of $\xi$, the corresponding curves are practically indistinguishable. 
}\label{fig:grey-body}
\end{figure*}

\begin{figure*}
\resizebox{0.9 \linewidth}{!}{\includegraphics{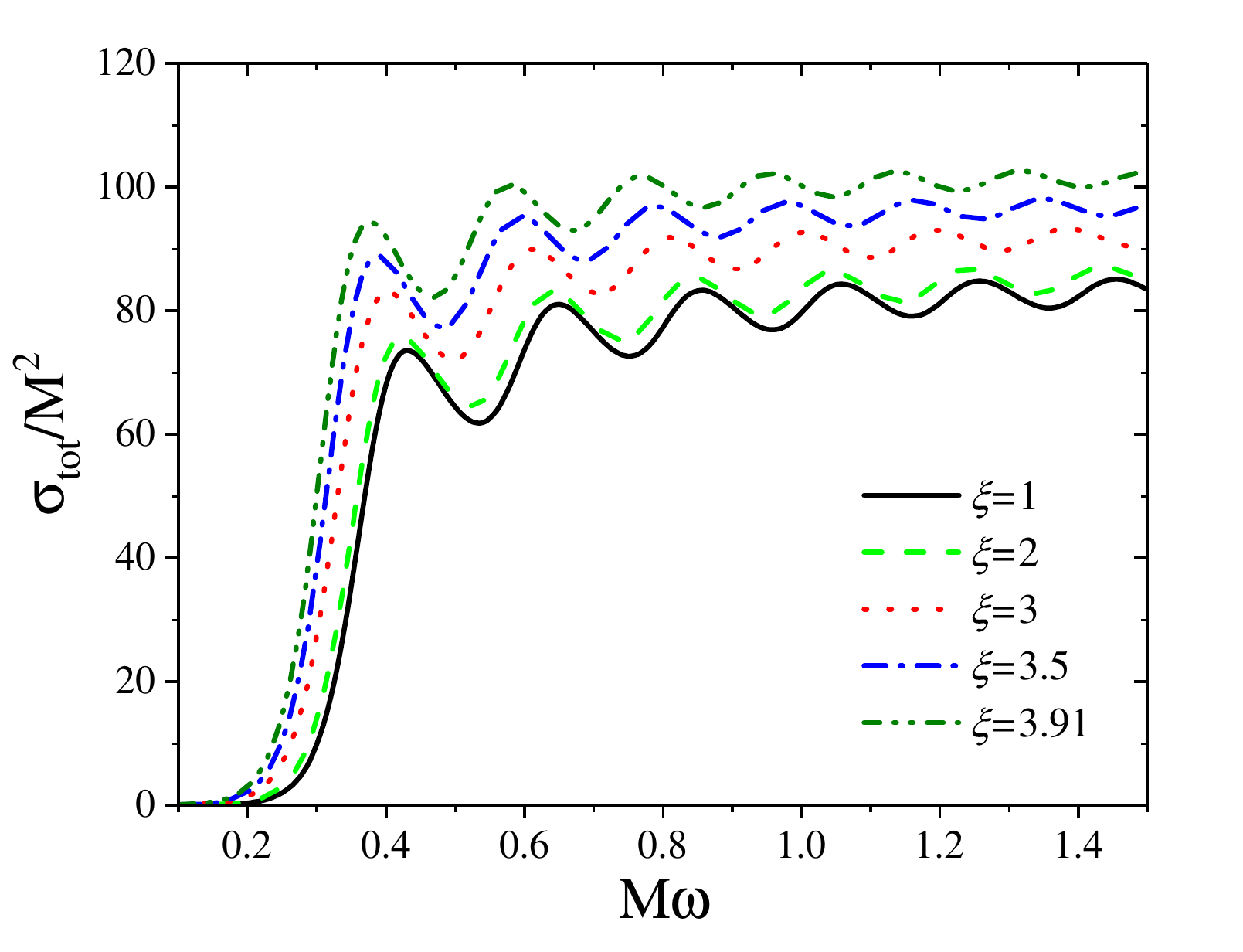}~~~\includegraphics{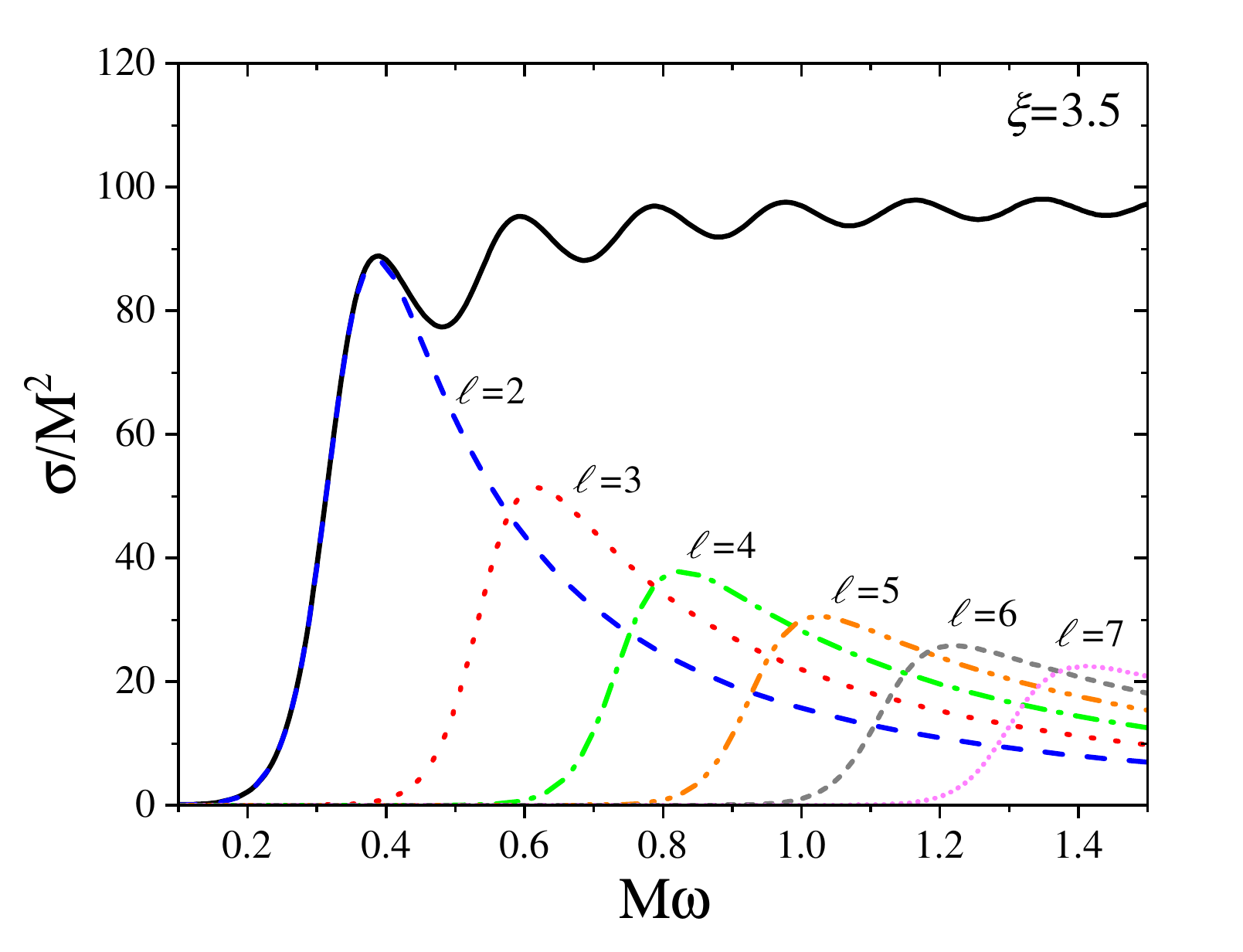}}
\caption{The typical behavior of       
         the total absorption cross-section for different values of  $\xi$ in the BH case (left panel).  For the lower values of $\xi$, the corresponding curves are practically indistinguishable. The contributions of the partial cross-sections with different $\ell$ to the total cross-section for  $\xi=3.5$ (right panel). 
}\label{fig:crossection}
\end{figure*}

\begin{figure}
\resizebox{0.9 \linewidth}{!}{\includegraphics{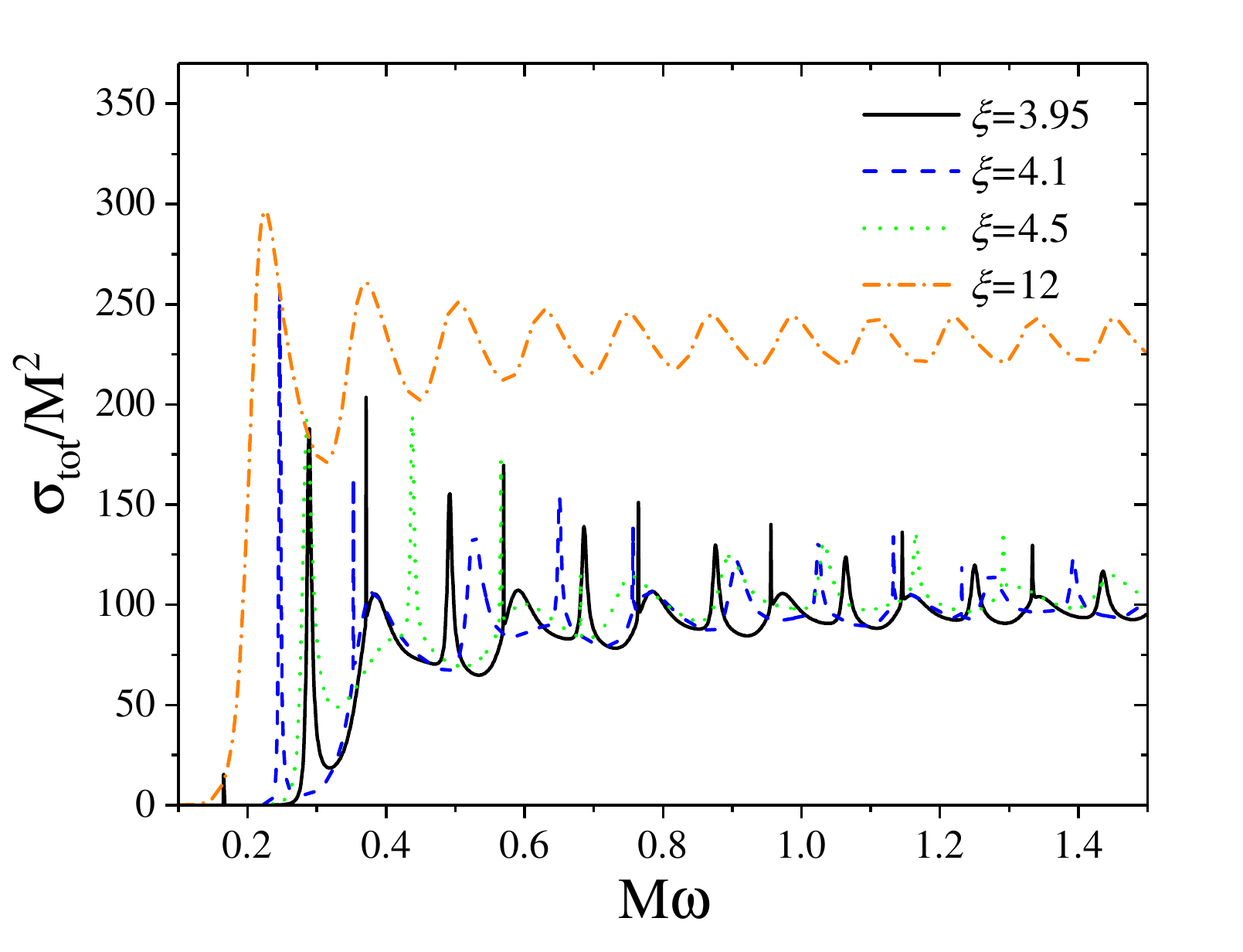}}
\caption{The typical behavior of the total absorption cross-section for different values of  $\xi$ in the WH case. The narrow peaks correspond to the quasi-resonances at specific values of $\omega$. 
}\label{fig:crossection:WH}
\end{figure}
\begin{figure*}
\resizebox{0.9 \linewidth}{!}{\includegraphics{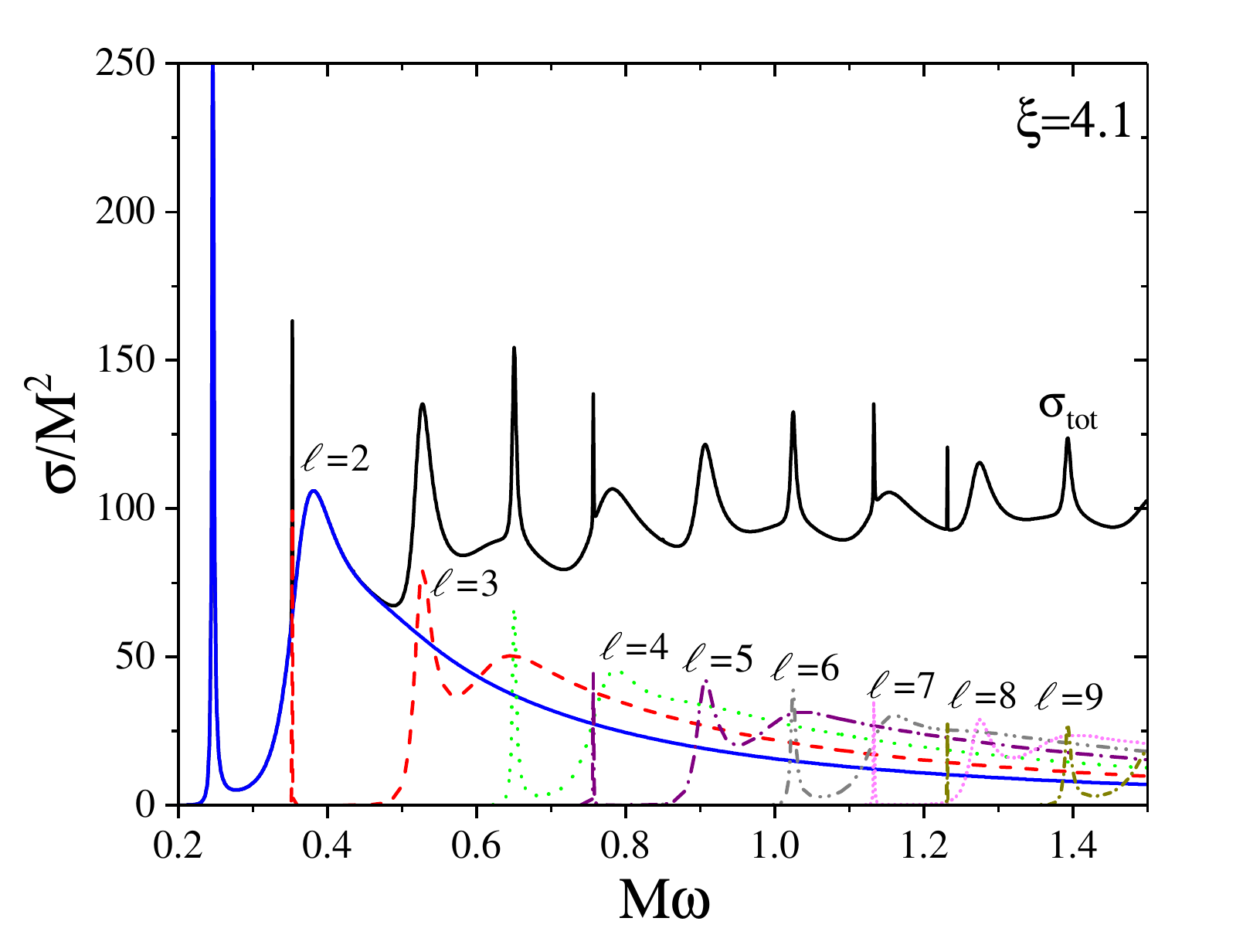}~~~\includegraphics{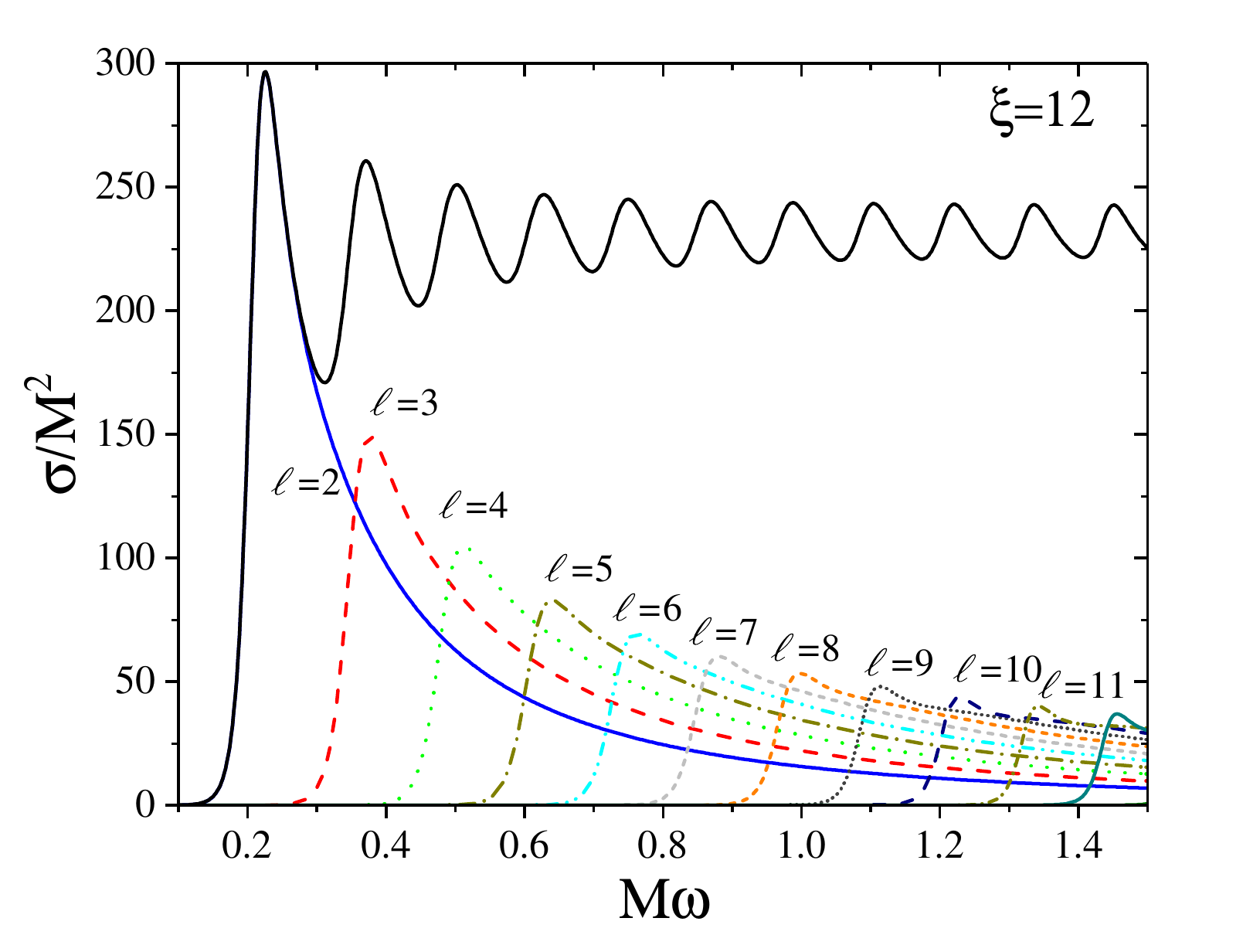}}
\caption{ The contributions of the partial cross-sections with different $\ell$ to the total cross-section for  $\xi=4.1$ (left) and $\xi=12$ (right).}\label{fig:crossectionWH_part}
\end{figure*}

\section{Scattering problems}

It is interesting to study properties of 
the monochromatic wave scattering on the black hole or wormhole background. For the scattering problem $\omega$ is real and the boundary conditions have the following form
\begin{equation}
\label{eq:BC_scatt}
\Psi = 
\begin{cases} 
e^{-i \omega r_*} + \mathcal{R}(\omega) e^{i \omega r_*}, & r_* \to \infty, \\
\mathcal{T}(\omega) e^{-i \omega r_*}, & r_* \to -\infty.
\end{cases}
\end{equation}
where $\mathcal{R}$ and $\mathcal{T}$ correspond to the reflection and transmission coefficients, respectively. They satisfy the following relation
\begin{equation}\label{1}
\left|\mathcal{T}\right|^2 + \left|\mathcal{R}\right|^2 = 1.
\end{equation}

These boundary conditions represent the process of wave scattering, where waves sent from  past null infinity in the first universe pass through the wormhole’s throat to the second universe. Part of the radiation is reflected by the effective potential, while some amount is transmitted to future null infinity in the second universe.
A similar scenario occurs in the black hole case, except that we now assume only an ingoing wave at the future black hole horizon.

While overtones of the quasinormal spectrum are highly sensitive to small deformations of the background spacetime \cite{Konoplya:2022pbc,Jaramillo:2020tuu}, it has been observed that grey-body factors are much more stable to such deformations \cite{Rosato:2024arw,Oshita:2024fzf} and could potentially be  used as GW observables \cite{Oshita:2023cjz} .
The grey-body factor $A_l$ is defined as  the transmission coefficient.
\begin{equation}
A_l\equiv \left|\mathcal{T}\right|^2=1-\left|\mathcal{R}\right|^2.
\end{equation}
In the BH case the use of grey-body factors is necessary ingredient for the estimation of the intensity of Hawking radiation. The quasi-thermal energy emission rate from the black hole horizon has the following form
\eq{
\frac{dE}{d\omega dt}=\frac{1}{\pi}\sum\limits_{\ell=2}^{\infty}(2\ell+1)A_l(\omega)\frac{\omega}{e^{\omega/T_h}-1},
}
where $T_h$ is the Hawking temperature, which is defined as follows 
\eq{
T_h=\frac{\sqrt{\mu(r)}f'(r)}{4\pi},\quad r=r_h,
}

The radiation phenomena around the compact objects
occurs in some volume. Therefore, such characteristics as absorption cross-section of a gravitational wave becomes important  \cite{Futterman:1988ni}.
We can estimate the effect of the absorbtion of a plane monochromatic wave by a black hole or a wormhole via calculating the absorbtion cross-sections. These have  been calculated for various black hole and wormhole models \cite{Azad:2020ajs,Lima:2020auu,Furuta:2024jpw,Li:2024xyu,Magalhaes:2023har,dePaula:2024xnd,PhysRevD.100.084001,Wu:2024ldo}.

The total absorption cross-section of the axial gravitational waves can be written as \cite{Futterman:1988ni}
\eq{
\sigma_{\rm tot}(\omega)=\frac{\pi}{\omega^2}\sum\limits_{\ell=2}^{\infty}(2\ell+1)A_l(\omega).
}
\subsection{Black hole}

To compute the grey-body factors in the BH case, one can use either the higher-order WKB approximation technique \cite{Konoplya:2019hlu} or direct numerical integration. We found that at small $\ell$ the WKB approximation provides good accuracy only for small values of $\xi/M < 1.5$. For larger values of $\xi$, the WKB results at various orders differ a lot and deviate considerably from the precise data. Consequently, to calculate transmission coefficients we  numerically solve (\ref{eq:master_eq_freq}) with boundary conditions (\ref{eq:BC_scatt}). Typical plots of the grey-body factors for black holes are shown in Fig. \ref{fig:grey-body}. The total and
partial absorption cross sections related to greybody factors
are shown in Fig.  \ref{fig:crossectionBH}. These plots reveal that the grey-body factors vary significantly depending on the value of the quantum parameter $\xi$.
The transmission coefficient is a monotonically increasing function of \(\omega\) for all values of \(\xi\). 
This can be easily explained by the shape of the effective potentials, which become lower for larger values of the quantum parameter \( \xi \), allowing a greater fraction of the flux to penetrate the barrier.

The intensity of Hawking radiation depends on two factors: the grey-body factors and the Hawking temperature of the event horizon. Typically, the temperature is the dominant factor determining the order of the energy emission rate. However, there are cases where the grey-body factors also play a significant role \cite{Bolokhov:2023ozp,Konoplya:2019ppy}. 
As shown in Fig. \ref{fig:Hawk_temp}, the temperature increases up to a certain large value of \( \xi \) and then the black hole begins to cool down. Considering that the grey-body factors also contribute to enhancing the radiation flux at larger \( \xi \), the total energy emission rate increases up to a value of \( \xi \) that exceeds the one at which the maximum Hawking temperature occurs.

The energy emission rate for the Hawking radiation is shown in Fig. \ref{fig:Hawking_radiation}. The dependence of the BH temperature $T_h$ on the quantum parameter $\xi$ is shown in fig. \ref{fig:Hawk_temp}.
The behaviour of total crosssection and influence of total crossections on total  are shown in Fig. \ref{fig:crossectionWH_part}.

\subsection{Wormhole}
In the wormhole case, however, the behavior of the transmission coefficient is significantly different. Due to the presence of a second barrier in the other universe, \(A_l\) becomes a non-monotonic function of \(\omega\), as illustrated in Fig.~\ref{fig:GB_WH}. The total and partial absorption cross sections are
shown in Figs. \ref{fig:crossection:WH} -\ref{fig:crossectionWH_part}  For small frequencies, multiple scattering can be observed, accompanied by the appearance of extremely narrow quasi-resonances at specific values of  \(\omega\).
In this context, it is compelling to test the correspondence between grey-body factors and quasinormal modes established in \cite{Konoplya:2024lir} for black holes and in \cite{Bolokhov:2024otn} for wormholes. This correspondence has been investigated in several recent studies \cite{Malik:2024cgb,Skvortsova:2024msa,Dubinsky:2024vbn}. From the precise numerical data obtained here for quasinormal modes and grey-body factors, we confirm that, if the multipole number $\ell$ is not small, the correspondence holds with high accuracy for black holes whenever the quantum parameter $\xi$ is not near its threshold value at the transition to the wormhole state
\cite{Konoplya:2024lir}:
\begin{equation}\label{transmission-eikonal}
\Gamma_{\ell}(\omega)=\left(1+e^{2\pi\dfrac{\omega^2-\re{\omega_0}^2}{4\re{\omega_0}\im{\omega_0}}}\right)^{-1} + \Sigma(\omega_{0},\omega_{1}),\\
\end{equation}
where $\Sigma(\omega_{0},\omega_{1})$ is the sum of the correction terms beyond the eikonal limit found in \cite{Konoplya:2024lir}. However, for wormholes considered in our case, the correspondence breaks down. This discrepancy can be readily explained by the form of the effective potential for wormholes, which features two peaks. This structure renders the standard WKB formula, used to derive the correspondence, inapplicable. This observation is consistent with \cite{Skvortsova:2024msa}, where wormholes with a single peak were studied.

\section{Conclusions}

In this work, we have explored the quasinormal modes and grey-body factors of quantum-corrected spacetimes, which describe either regular black holes or traversable wormholes depending on the value of the quantum parameter $\xi$. Our results reveal several key findings:

\begin{itemize}
    \item Deviation of overtones: For the black hole state, while the fundamental quasinormal mode changes only mildly (by approximately 10-15\%) compared to the Schwarzschild solution, the higher overtones exhibit significantly stronger deviations, exceeding 30\% for the third overtone. These deviations grow with the overtone number and are linked to the near-horizon modifications of the spacetime geometry.

    \item Transition to wormholes: The transition to the wormhole state, marked by $\xi > 3.93$, leads to a drastic change in the spectrum of quasinormal modes. The wormhole state is characterized by long-lived modes, and the time-domain evolution shows the presence of echoes in the early ringdown phase, followed by a regular ringdown phase dominated by slowly decaying modes. These long-lived modes are non-perturbative in $\xi$ and cannot be smoothly connected to the Schwarzschild-like modes of the black hole.

    \item Grey-body factors:
    The grey-body factors of black holes and wormholes depend significantly on the quantum parameter $\xi$. For black holes, the correspondence between quasinormal modes and grey-body factors holds accurately for higher multipole numbers $\ell$. However, for wormholes, this correspondence breaks down due to the double-peaked structure of the effective potential, which is inconsistent with the assumptions of the standard WKB formula.

    \item Long-lived modes:
    As $\xi$ asymptotically approaches the critical threshold value of the transition, the damping rate of the wormhole modes decreases, giving rise to arbitrarily long-lived modes. These modes bear similarity to those observed in spacetimes with massive field perturbations and could potentially contribute to low-frequency gravitational waves detectable by Pulsar Timing Array experiments.
\end{itemize}

Our results demonstrate that quantum corrections introduce distinct features in the quasinormal spectra and grey-body factors of black holes and wormholes. These results provide a pathway to observationally distinguish quantum-corrected spacetimes from classical black holes and wormholes. Future work could focus on extending this analysis to rotating spacetimes and exploring potential observational signatures in the context of gravitational wave astronomy.

\begin{acknowledgments}
R. A. K.  acknowledges A. Zhidenko for useful discussions. O.S. is supported in part by National Science 
Foundation grant PHY-2110466 and the Tufts Scholar at Risk Program.
\end{acknowledgments}
\FloatBarrier
\appendix
\section{Tables with accurate values of the quasinormal modes }
\label{sec:appendix}
\begin{table}
\begin{tabular}{c| c| c| c|}
\hline
\hline
$\xi $ & $n$ & $\ell=2$ & $\ell=3$ \\
\hline
$0.01$ & \makecell{$0$ \\ $1$ \\ $2$ \\ $3$}&\makecell{$0.373672 - \
0.088962i$ \\ $0.346711 - 0.273915i$ \\ $0.301053 - 0.478277i$ \\ 
$0.251505 - 0.705148i$} & \makecell{$0.599443 - 0.092703i$ \\ 
$0.582644 - 0.281298i$ \\ $0.551685 - 0.479093i$ \\ $0.511962 - \
0.690337i$}\\
 \hline $0.1$ & \makecell{$0$ \\ $1$ \\ $2$ \\ 
$3$}&\makecell{$0.373672 - 0.088962i$ \\ $0.346711 - 0.273915i$ \\ 
$0.301053 - 0.478277i$ \\ $0.251505 - 0.705148i$} & \
\makecell{$0.599443 - 0.092703i$ \\ $0.582644 - 0.281298i$ \\ 
$0.551685 - 0.479093i$ \\ $0.511962 - 0.690337i$}\\
 \hline $0.5$ & \makecell{$0$ \\ $1$ \\ $2$ \\ 
$3$}&\makecell{$0.373632 - 0.088987i$ \\ $0.346651 - 0.273991i$ \\ 
$0.300995 - 0.478412i$ \\ $0.251563 - 0.705323i$} & \
\makecell{$0.599397 - 0.092729i$ \\ $0.582588 - 0.281381i$ \\ 
$0.551620 - 0.479244i$ \\ $0.511905 - 0.690575i$}\\
 \hline $1.$ & \makecell{$0$ \\ $1$ \\ $2$ \\ 
$3$}&\makecell{$0.373035 - 0.089340i$ \\ $0.345745 - 0.275082i$ \\ 
$0.300061 - 0.480333i$ \\ $0.252199 - 0.707929i$} & \
\makecell{$0.598702 - 0.093109i$ \\ $0.581747 - 0.282579i$ \\ 
$0.550598 - 0.481436i$ \\ $0.510922 - 0.694008i$}\\
 \hline $2.$ & \makecell{$0$ \\ $1$ \\ $2$ \\ 
$3$}&\makecell{$0.364591 - 0.092566i$ \\ $0.332508 - 0.284386i$ \\ 
$0.283054 - 0.496829i$ \\ $0.240030 - 0.733901i$} & \
\makecell{$0.588674 - 0.097139i$ \\ $0.568312 - 0.294880i$ \\ 
$0.531469 - 0.502883i$ \\ $0.486170 - 0.726263i$}\\
 \hline 
 $2.5$ & \makecell{$0$ \\ $1$ \\ $2$ \\ $3$}&\makecell{$0.355063 - \
0.094058i$ \\ $0.318803 - 0.288440i$ \\ $0.262491 - 0.504804i$ \\ 
$0.214032 - 0.741562i$} & \makecell{$0.576728 - 0.099868i$ \\ 
$0.552164 - 0.302846i$ \\ $0.507530 - 0.515575i$ \\ $0.452822 - \
0.742816i$}\\
 \hline $3.5$ & \makecell{$0$ \\ $1$ \\ $2$ \\ 
$3$}&\makecell{$0.330595 - 0.093030i$ \\ $0.289794 - 0.288070i$ \\ 
$0.206291 - 0.504311i$ \\ $0.16874 - 0.70576i$} & \makecell{$0.543564 \
- 0.102021i$ \\ $0.511358 - 0.309672i$ \\ $0.447554 - 0.525527i$ \\ 
$0.369838 - 0.741451i$}\\
 \hline $3.8$ & \makecell{$0$ \\ $1$ \\ $2$ \\ 
$3$}&\makecell{$0.322995 - 0.091843i$ \\ $0.282121 - 0.286363i$ \\ 
$0.187355 - 0.499030i$ \\ $0.16473 - 0.68955i$} & \makecell{$0.532699 \
- 0.101713i$ \\ $0.499126 - 0.309276i$ \\ $0.429579 - 0.525004i$ \\ 
$0.346600 - 0.732100i$}\\
 \hline 
\end{tabular}
\label{tab:QNMs_BH}
\caption{The values of fundamental quasinormal mode and first three overtones for different values of $\xi$, for $\ell=2$ and $\ell=3$ in the black hole case. }
\end{table}

\begin{table}
\begin{tabular}{c| c| c| c|}
\hline
\hline
$\xi $ & $n$ & $\ell=2$ & $\ell=3$ \\
\hline
$3.94$ & \makecell{$0$ \\ $1$ \\ $2$ \\ $3$ \\ $4$}&\makecell{$        
0.1330456 - 5.7\cdot10^{-6}  i$ \\ $0.244940 - 0.000800i$ \\ $0.330575 - 0.010333i$ \\ $0.403691 - 0.039308i$ \\ $0.484038 - 0.082009i$} & \makecell{$           
0.160204 - 6.6\cdot10^{-9}  i$ \\ $ 
0.297943 - 5.23 \cdot10^{-6}  i$ \\ $0.413628 - 0.000344i$ \\ $0.508062 - 0.005347i$ \\ $0.586486 - 0.026292i$}\\
 \hline $4.$ & \makecell{$0$ \\ $1$ \\ $2$ \\ $3$ \\ $4$}&\makecell{$0.212791 - 0.000391i$ \\ $0.337370 - 0.016978i$ \\ $0.421615 - 0.075165i$ \\ $0.524901 - 0.168412i$ \\ $0.659149 - 0.272118i$} & \makecell{$ 
0.288562 - 6.83\cdot 10^{-6}  i$ \\ $0.466434 - 0.002684i$ \\ $0.574309 - 0.031294i$ \\ $0.662790 - 0.103992i$ \\ $0.765095 - 0.203750i$}\\
 \hline $4.3$ & \makecell{$0$ \\ $1$ \\ $2$ \\ $3$ \\ $4$}&\makecell{$0.272626 - 0.004991i$ \\ $0.380805 - 0.060541i$ \\ $0.461223 - 0.188904i$ \\ $0.590437 - 0.376044i$ \\ $0.178879 - 0.528720i$} & \makecell{$0.410004 - 0.001022i$ \\ $0.561213 - 0.037907i$ \\ $0.639072 - 0.140820i$ \\ $0.730651 - 0.299223i$ \\ $0.846490 - 0.499214i$}\\
 \hline $4.5$ & \makecell{$0$ \\ $1$ \\ $2$ \\ $3$ \\ $4$}&\makecell{$0.283858 - 0.008343i$ \\ $0.384245 - 0.078065i$ \\ $0.459838 - 0.232290i$ \\ $0.590771 - 0.459387i$ \\ $0.164329 - 0.509905i$} & \makecell{$0.437423 - 0.003001i$ \\ $0.573143 - 0.058422i$ \\ $0.640710 - 0.186470i$ \\ $0.725360 - 0.373128i$ \\ $0.841447 - 0.618468i$}\\
 \hline $4.8$ & \makecell{$0$ \\ $1$ \\ $2$ \\ $3$ \\ $4$}&\makecell{$0.290591 - 0.012726i$ \\ $0.381709 - 0.097035i$ \\ $0.448131 - 0.279562i$ \\ $0.144295 - 0.484358i$ \\ $0.144408 - 0.493452i$} & \makecell{$0.457029 - 0.007081i$ \\ $0.575335 - 0.082556i$ \\ $0.630966 - 0.235485i$ \\ $0.700508 - 0.454685i$ \\ $0.356970 - 0.657491i$}\\
 \hline $5.$ & \makecell{$0$ \\ $1$ \\ $2$ \\ $3$ \\ $4$}&\makecell{$0.291838 - 0.015176i$ \\ $0.377798 - 0.106579i$ \\ $0.437552 - 0.303736i$ \\ $0.132216 - 0.469234i$ \\ $0.134691 - 0.480411i$} & \makecell{$0.463009 - 0.010003i$ \\ $0.572116 - 0.095253i$ \\ $0.620891 - 0.259848i$ \\ $0.679441 - 0.497243i$ \\ $0.336059 - 0.640180i$}\\
 \hline $6.$ & \makecell{$0$ \\ $1$ \\ $2$ \\ $3$ \\ $4$}&\makecell{$0.284241 - 0.023153i$ \\ $0.351278 - 0.134607i$ \\ $0.379068 - 0.378657i$ \\ $0.086784 - 0.410481i$ \\ $0.101553 - 0.424028i$} & \makecell{$0.461453 - 0.022286i$ \\ $0.539238 - 0.134435i$ \\ $0.560154 - 0.330087i$ \\ $0.277912 - 0.573587i$ \\ $0.257981 - 0.573809i$}\\
 \hline $8.$ & \makecell{$0$ \\ $1$ \\ $2$ \\ $3$ \\ $4$}&\makecell{$0.255811 - 0.028675i$ \\ $0.301622 - 0.150995i$ \\ $0.040605 - 0.336145i$ \\ $0.066098 - 0.341606i$ \\ $0.289291 - 0.432580i$} & \makecell{$0.422539 - 0.034128i$ \\ $0.414231 - 0.163975i$ \\ $0.453835 - 0.370338i$ \\ $0.144451 - 0.400996i$ \\ $0.176686 - 0.485606i$}\\
 \hline $10.$ & \makecell{$0$ \\ $1$ \\ $2$ \\ $3$ \\ $4$}&\makecell{$0.230363 - 0.029507i$ \\ $0.264524 - 0.151202i$ \\ $0.045120 - 0.288070i$ \\ $0.018921 - 0.288274i$ \\ $0.234013 - 0.440000i$} & \makecell{$0.383433 - 0.037876i$ \\ $0.414231 - 0.163975i$ \\ $0.378644 - 0.371104i$ \\ $0.144451 - 0.400996i$ \\ $0.135426 - 0.425350i$}\\
 \hline $15.$ & \makecell{$0$ \\ $1$ \\ $2$ \\ $3$ \\ $4$}&\makecell{$0.185648 - 0.027392i$ \\ $0.205702 - 0.138252i$ \\ $0.020177 - 0.213618i$ \\ $0.159818 - 0.409919i$ \\ $0.197513 - 0.753587i$} & \makecell{$0.311646 - 0.037805i$ \\ $0.325366 - 0.152601i$ \\ $0.087575 - 0.306743i$ \\ $0.087793 - 0.331538i$ \\ $0.267938 - 0.340932i$}\\
 \hline $100.$ & \makecell{$0$ \\ $1$ \\ $2$ \\ $3$ \\ $4$}&\makecell{$0.057804 - 0.010402i$ \\ $0.059503 - 0.053331i$ \\ $0.030836 - 0.163177i$ \\ $0.030081 - 0.292194i$ \\ $0.033496 - 0.417437i$} & \makecell{$0.098272 - 0.015704i$ \\ $0.096249 - 0.059744i$ \\ $0.011789 - 0.087259i$ \\ $0.017446 - 0.095086i$ \\ $0.055777 - 0.137056i$}\\
 \hline \hline
\end{tabular}
\caption{The values of fundamental quasinormal mode and first four overtones for different values of $\xi$, for $\ell=2$ and $\ell=3$ in the wormhole case.  }
\label{tab:QNMs_WH}
\end{table}

\begin{table}
\centering
\begin{tabular}{|c|c|c|}
\hline
\hline
$\xi$ & QNM ($\ell = 0$) & QNM ($\ell = 1$) \\
\hline
0.1 & $0.110751 - 0.105068 i$ & $0.292933 - 0.097658 i$ \\
1 &   $0.110080 - 0.105896 i$ & $0.292656 - 0.098345 i$ \\
1.5 & $0.107774 - 0.109711 i$ & $0.291397 - 0.100759 i$ \\
2 &   $0.101472 - 0.114984 i$ & $0.287831-  0.105289 i$ \\
2.5 & $0.090930 - 0.115453 i$ & $0.281374 - 0.110588 i$ \\
3 &   $0.080501 - 0.116241 i$ & $0.272822 - 0.115221 i$ \\
3.5 & $0.070456 - 0.115160 i$ & $0.263319 - 0.118651i$ \\
%3.82 & $-$  & $0.531974 - 0.101679 i$ \\
3.9 & $0.063469 - 0.113837 i$ & $0.255590 - 0.120549 i$ \\
\hline
4.0 & $0.127333 - 0.024379 i$ & $0.196274 - 0.003086 i$ \\
4.2 & $0.148266 - 0.046168 i$ & $0.243321 - 0.013192 i$ \\
4.4 & $0.154633 - 0.059822 i$ & $0.260889 - 0.022352 i$ \\
4.6 & $0.156867 - 0.070305 i$ & $0.269891 - 0.030333 i$ \\
4.8 & $0.157050 - 0.077978 i$ & $0.274679 - 0.037253 i$ \\
5.0 & $0.156854 - 0.085256 i$ & $0.276999 - 0.043257 i$ \\
\hline
\hline
\end{tabular}
\caption{The fundamental quasinormal modes of a scalar field perturbations obtained by the time-domain integration for different values of $\xi$, $M=1$, $\ell=0$ and $1$. At the wormhole transition $\xi \approx 3.93$ the late-time dominant frequency changes drastically.}
\label{tab:QNMs_WHScalar}
\end{table}

\begin{table}
\centering
\begin{tabular}{|c|c|c|}
\hline
\hline
$\xi$ & QNM ($\ell = 1$) & QNM ($\ell = 2$) \\
\hline
0.1 & $0.248268 - 0.092489 i$ & $0.457596 - 0.095004 i$ \\
1 &   $0.247768 - 0.093000 i$ & $0.457082 - 0.095495 i$ \\
1.5 & $0.245785 - 0.094685 i$ & $0.454997 - 0.097227 i$ \\
2 &   $0.241122 - 0.097418 i$ & $0.449762 - 0.100455 i$ \\
2.5 & $0.233859 - 0.099892 i$ & $ 0.44080 - 0.104064 i$ \\
3 &   $0.225172 - 0.101218 i$ & $0.429038 - 0.106838 i$ \\
3.5 & $0.216148 - 0.101340 i$ & $0.216148 - 0.101340 i$ \\
%3.82 & $-$  & $0.531974 - 0.101679 i$ \\
3.9 & $0.209111 - 0.100756 i$ & $0.209111 - 0.100756 i$ \\
\hline
4.0 & $0.175947 - 0.004908 i$ & $0.251764 - 0.000191 i$ \\
4.2 & $0.209061 - 0.015832 i$ & $0.326691 - 0.003010 i$ \\
4.4 & $0.219369 - 0.023883 i$ & $0.354516 - 0.007757 i$ \\
4.6 & $0.223496 - 0.030154 i$ & $0.368686 - 0.012952 i$ \\
4.8 & $0.224760 - 0.035169 i$ & $0.376323 - 0.017976 i$ \\
5.0 & $0.224447 - 0.039240 i$ & $0.380216 - 0.022609 i$ \\
\hline
\hline
\end{tabular}
\caption{The fundamental quasinormal modes of an electromagnetic perturbations obtained by the time-domain integration for different values of $\xi$, $M=1$, $\ell=0$ and $1$. At the wormhole transition $\xi \approx 3.93$ the late-time dominant frequency changes drastically.}
\label{tab:QNMs_WHElmag}
\end{table}

\begin{table}
\centering
\begin{tabular}{|c|c|c|}
\hline
\hline
$\xi$ & QNM ($\ell = 2$) & QNM ($\ell = 3$) \\
\hline
0.1 & $0.373672 - 0.088962 i$ & $0.599445 - 0.092702 i$ \\
1 & $0.373035 - 0.089339 i$ & $0.598703 - 0.093109 i$ \\
1.5 & $0.370530 - 0.090589 i$ & $0.595777 - 0.094533 i$ \\
2 & $0.364591 - 0.092566 i$ & $0.588676 - 0.097138 i$ \\
2.5 & $0.355063 - 0.094058 i$ & $0.576729 - 0.099868 i$ \\
3 & $0.343254 - 0.094208 i$ & $0.561111 - 0.101603 i$ \\
3.5 & $0.330595 - 0.093029 i$ & $0.543565 - 0.102021 i$ \\
3.82 & $0.322493 - 0.091753 i$ & $0.531974 - 0.101679 i$ \\
3.9 &  $0.320490 - 0.091387 i$ & $0.529071 - 0.101532 i$ \\
\hline
4.0 & $0.212791 - 0.000391 i$ & $0.288562 - 6.834\cdot10^{-6} i$ \\
4.2 & $0.262580 - 0.003274 i$ & $0.387304 - 0.000422 i$ \\
4.4 & $0.279288 - 0.006698 i$ & $0.425896 - 0.001898 i$ \\
4.6 & $0.287020 - 0.009900 i$ & $0.287020 - 0.009900 i$ \\
4.8 & $0.290591 - 0.012726 i$ & $0.457029 - 0.007081 i$ \\
5.0 & $0.291838 - 0.015176 i$ & $0.463010 - 0.010003 i$ \\
\hline
\hline
\end{tabular}
\caption{The fundamental quasinormal modes obtained by the time-domain integration for different values of $\xi$, $M=1$, $\ell=2$ and $3$. At the wormhole transition $\xi \approx 3.93$ the late-time dominant frequency changes drastically.}
\label{tab:QNMs_AxBHWHtd}
\end{table}

\FloatBarrier
%\newpage
%\bibliographystyle{unsrt}
\bibliography{bibliography}
\end{document}